\documentclass[fleqn,usenatbib]{mnras}

\usepackage{newtxtext,newtxmath}
\usepackage[T1]{fontenc}

\DeclareRobustCommand{\VAN}[3]{#2}
\let\VANthebibliography\thebibliography
\def\thebibliography{\DeclareRobustCommand{\VAN}[3]{##3}\VANthebibliography}

\usepackage{graphicx}	% Including figure files
\usepackage{amsmath}	% Advanced maths commands
\usepackage{booktabs}
\usepackage{caption}
\usepackage{subcaption}

\title[Wide PCEBs with ultramassive WDs]{Wide post-common envelope binaries containing ultramassive white dwarfs: evidence for efficient envelope ejection in massive AGB stars}

\author[]{Natsuko Yamaguchi$^{1}$\thanks{E-mail: nyamaguc@caltech.edu},
Kareem El-Badry$^{1}$, Jim Fuller$^{1}$, David W. Latham$^{2}$, Phillip A. Cargile$^{2}$, \newauthor  Tsevi Mazeh$^{3}$, Sahar Shahaf$^{4}$, Allyson Bieryla$^{2}$, Lars A. Buchhave$^{6}$, Melissa Hobson$^{5}$
\\
$^{1}$ Department of Astronomy, California Institute of Technology, 1200 E. California Blvd, Pasadena, CA, 91125, USA \\
$^{2}$ Center for Astrophysics $|$ Harvard \& Smithsonian, 60 Garden Street, Cambridge, MA 02138, USA \\
$^{3}$ School of Physics and Astronomy, Tel Aviv University, Tel Aviv, 6997801, Israel \\
$^{4}$ Department of Particle Physics and Astrophysics, Weizmann Institute of Science, Rehovot 7610001, Israel \\
$^{6}$ DTU Space, National Space Institute, Technical University of Denmark, Elektrovej 328, DK-2800 Kgs. Lyngby, Denmark \\
$^{5}$ Max-Planck-Institute for Astronomy, Königstuhl 17, 69117 Heidelberg, Germany 
}

\date{Accepted XXX. Received YYY; in original form ZZZ}

\pubyear{2015}

\begin{document}
\label{firstpage}
\pagerange{\pageref{firstpage}--\pageref{lastpage}}
\maketitle

\begin{abstract}
% proposed rewording (will tweak a bit more) 

Post-common-envelope binaries (PCEBs) containing a white dwarf (WD) and a main-sequence (MS) star can constrain the physics of common envelope evolution and calibrate binary evolution models. Most PCEBs studied to date have short orbital periods ($P_{\rm orb} \lesssim 1$\,d), implying relatively inefficient harnessing of binaries' orbital energy for envelope expulsion. Here, we present follow-up observations of five binaries from {\it Gaia} DR3 containing solar-type MS stars and probable ultramassive WDs ($M\gtrsim 1.2\,M_{\odot}$) with significantly wider orbits than previously known PCEBs, $P_{\rm orb} = 18-49$\,d. The WD masses are much higher than expected for systems formed via stable mass transfer at these periods, and their near-circular orbits suggest partial tidal circularization when the WD progenitors were giants. These properties strongly suggest that the binaries are PCEBs. Forming PCEBs at such wide separations requires highly efficient envelope ejection, and we find that the observed periods can only be explained if a significant fraction of the energy released when the envelope recombines goes into ejecting it. Our 1D stellar models including recombination energy confirm prior predictions that a wide range of PCEB orbital periods, extending up to months or years, can potentially result from Roche lobe overflow of a luminous AGB star. This evolutionary scenario may also explain the formation of several wide WD+MS binaries discovered via self-lensing, as well as a significant fraction of post-AGB binaries and barium stars. 

\end{abstract}


%%%%%%%%%%%%%%%%%%%%%%%%%%%%%%%%%%%%%%%%%%%%%%%%%%

%%%%%%%%%%%%%%%%% BODY OF PAPER %%%%%%%%%%%%%%%%%%

\renewcommand{\arraystretch}{1.1}

\section{Introduction}

Common envelope evolution (CEE) is a major unsolved problem in binary evolution. CEE is the outcome of dynamically unstable mass transfer (MT), generally from a more massive donor to a less massive accretor. During CEE, both stars orbit inside a shared envelope, spiraling inward on a dynamical or thermal timescale. In some cases, the orbital energy liberated during this inspiral is sufficient to eject the shared envelope, leaving behind a close binary in which at least one component has lost most of its envelope. If envelope ejection is {\it not} successful, the final outcome of CEE is a stellar merger. Modeling of CEE is a key uncertainty in our understanding of the formation of a wide variety of binary systems, including cataclysmic variables \citep[e.g.][]{Paczynski1976, Meyer1979, Willems2004}, X-ray binaries \citep[e.g.][]{Kalogera1998ApJ},  Type Ia supernovae \citep[e.g.][]{Webbink1984, Meng2017MNRAS}, binary neutron stars \citep[e.g.][]{Bhattacharya1991}, and binary black holes \citep{Belczynski2016, Marchant2021A&A}. 

CEE is a dynamical process often involving an enormous range of physical and temporal scales. Because detailed, end-to-end calculations of CEE are currently infeasible \citep[e.g.][]{Ivanova2013A&ARv} -- and because it is often necessary to model the evolution of large numbers of binaries to understand the possible formation pathways of a single observed system -- binary population synthesis \citep[BPS; e.g.][]{Hurley2002} codes are often used to model the evolution of millions of binaries, making it possible to explore a broad parameter space at the expense of physical realism. These codes make use of simplified models of CEE based on energy or angular momentum conservation. In the most widely used formalism, it is assumed that a fixed fraction (``$\alpha$'') of the liberated orbital energy goes in ejecting the envelope. This fraction, and the binding energy of the envelope, then sets the post-CEE orbital separation \citep[e.g.][]{Livio1988, Tout1997, DeMarco2011MNRAS}. Other energy sources, such as the photons released during recombination when the envelope expands, are often modeled as reducing the envelope's binding energy.

CEE models have historically been calibrated by comparing the binary populations they predict to observed post-common envelope binaries (PCEBs). The most abundantly observed PCEBs contain white dwarfs or hot subdwarfs in tight orbits ($P_{\rm orb} \lesssim 1$\,d) with a main-sequence (MS) star, having been produced when the MS star spiraled through the envelope of a red giant and ultimately ejected it. BPS models have succeeded in explaining some broad population properties of these binaries when using the $\alpha$-formalism \citep{Han2002, Han2003, Camacho2014A&A}. Such modeling makes it possible to empirically constrain $\alpha$, and several calculations have found that the observations can best be reproduced by models assuming $\alpha \approx 0.3$, meaning that $\approx 30\%$ of the orbital energy liberated during inspiral goes into ejecting the envelope  \citep{Zorotovic2010A&A, Davis2012, Toonen2013A&A, Camacho2014A&A, Zorotovic2022, Scherbak2023MNRAS}.

At least one PCEB is known with a (relatively) wide orbit. That system, IK\,Peg, has a period of 22 days and hosts an unusually massive WD, with mass $M\approx 1.2\,M_{\odot}$ \citep{Wonnacott1993}. The system's wide orbit ($a\approx 0.2$\,au) means that less orbital energy was liberated during the MS star's inspiral than in typical PCEBs with $a\approx 0.01$\,au. \citet{Zorotovic2010A&A} found that in the $\alpha-$formalism, IK\,Peg's orbit can {\it only} be explained when additional sources of energy besides orbital inspiral are taken into account. Besides IK\,Peg, several wide WD+MS binaries have been discovered via self-lensing, with orbital periods ranging from a few months to a few years  \citep{Kruse2014Sci, Kawahara2018AJ}. While it is not clear whether these systems formed via CEE (see Section \ref{ssec:lit_pcebs}), they are also candidates for being wide PCEBs and would require additional energy sources (and/or  high $\alpha$ values) to explain \citep{Zorotovic2014A&A}. Energy released by H and He recombination in the expanding envelope of the WD progenitor is a prime suspect for supplying the additional energy (first explored by \citealt{1968IAUS...34..396P}, and later studied by e.g. \citealt{Webbink2008ASSL, Ivanova2015MNRAS, Ivanova2018ApJL})

Most PCEBs studied to date were identified via their composite spectra and RV variability detectable with low-resolution spectra \citep{Rebassa-Mansergas2007MNRAS, Rebassa-Mansergas2017MNRAS, Lagos2022MNRAS}. This leads to strong selection effects in favor of PCEBs containing low-mass MS stars (which are less likely to outshine the WD) in tight orbits (where RV shifts are larger). The recent 3rd data release of the {\it Gaia} mission \citep[DR3;][]{GaiaCollaboration2023} contains orbital solutions for more than $10^5$ astrometric binaries, and for more than $10^5$ single-lined spectroscopic binaries identified from medium-resolution spectra \citep{GaiaCollaboration2023A&A}.
This dataset provides a new opportunity to search for PCEBs with wider orbits and more massive MS companions.

In this paper, we present five binaries in relatively wide orbits containing solar-type main sequence stars and probable ultramassive WD candiates. Section \ref{sec:discovery} describes our identification of wide PCEB candidates from the {\it Gaia} DR3 catalog. Section \ref{sec:follow-up} describes follow-up spectroscopic observations to obtain radial velocities (RVs), spectral analysis to calculate metallicities, and fitting to the broadband spectral energy distributions to constrain stellar parameters of the MS stars. In Section \ref{sec:rvs}, we fit the RVs to infer orbital solutions. In particular, we measure a mass function which, when combined with the luminous star mass, yields a minimum mass for the compact object. We also discuss alternative possibilities for the nature of the unseen companions. In Section \ref{sec:comparison}, we compare our systems to other known PCEBs. Section \ref{sec:mesa} describes models of the massive WD progenitors and constraints on CEE. In Section \ref{sec:discussion}, we briefly describe an alternative CE formalism, the occurrence rate of close and wide PCEBs, and selection biases in past surveys. Finally, in Section \ref{sec:conclusion}, we summarize our main results and conclude.

\section{Discovery} \label{sec:discovery}

The five objects studied in this paper were discovered in the course of a broader search for compact objects with single-lined spectroscopic (``SB1'') or astrometric + spectroscopic (``AstroSpectroSB1'') solutions in the {\it Gaia} DR3 non-single star (NSS) catalog \citep{GaiaCollaboration2023A&A}.
We selected promising candidates for further follow-up based on their mass functions, color-magnitude diagram (CMD) positions, and {\it Gaia} quality flags. In brief, we targeted sources whose CMD positions suggested a single luminous source and whose {\it Gaia} mass functions implied a companion mass near the Chandrasekhar limit. For objects with SB1 solutions, we prioritized those for which \cite{Bashi2022MNRAS} reported a robustness ``score'' above 0.5, corresponding roughly to an expected 20\% contamination rate with spurious solutions.

Our spectroscopic follow-up revealed some sources to have spurious {\it Gaia} orbital solutions and others to be double- or triple-lined binaries. Here, we focus on 5 promising sources that are single-lined and whose {\it Gaia}-reported orbits were validated by our follow-up. All 5 of these sources turned out to have near-circular orbits, but eccentricity did not enter our initial selection, and we did not find any similar (single-lined, high mass function) targets with comparable periods and higher eccentricities. The names, {\it Gaia} DR3 source IDs, and basic information of these 5 objects are summarized in Table \ref{tab:basic_info}. Our full search will be described in future work. 

Four targets have spectroscopic SB1 solutions, but no astrometric binary solution. We suspect this is a result of the stringent cuts on astrometric signal-to-noise ratio applied to astrometric solutions with short periods \citep[see][]{GaiaCollaboration2023A&A}. For these objects, the inclination is unknown, and only a minimum companion mass can be inferred. One object, J1314+3818, has a joint astrometric and spectroscopic (``AstroSpectroSB1'') solution, meaning that its inclination is constrained. This object was identified as a  likely MS + compact object binary by \citet{Shahaf2023MNRAS} on the basis of its large astrometric mass ratio function \citep[also see][]{Shahaf2023arXiv}. Another object in our sample, J2034-5037, was previously identified by \citet{Jayasinghe2023MNRAS} as a candidate neutron star + MS binary.

\begin{table*}
    \centering
    \begin{tabular}{c c c c c c c}
         \hline
         Name & Gaia DR3 ID & RA [deg] & Dec [deg] & $G$ [mag] & RUWE & $\varpi [\rm mas]$ \\
         \hline
         J2117+0332 & 2692960678029100800 & 319.34490 & 3.54044 & 12.47 & 1.27 & 1.96 $\pm$ 0.02 \\
         J1111+5515 & 843829411442724864 & 167.80947 & 55.26410 & 10.61 & 1.47 & 3.24 $\pm$ 0.02 \\
         J1314+3818 & 1522897482203494784 & 198.51734 & 38.30119 & 11.05 & - & 12.45 $\pm$ 0.02 \\
         J2034-5037 & 6475655404885617920 & 308.60840 & -50.62557 & 12.37 & 2.94 & 3.23 $\pm$ 0.04 \\
         J0107-2827 & 5033197892724532736 & 16.98021 & -28.46128 & 12.27 & 1.74 & 2.14 $\pm$ 0.02 \\
         \hline
    \end{tabular}
    \caption{Basic information from Gaia DR3 of the five objects found in this work. The format for the name of each object is `J' for J2000 followed by the coordinates of the right ascension (RA) in hours and minutes, and declination (Dec) in degrees and minutes. $G$ is the G-band mean magnitude, RUWE is the Renormalised Unit Weight Error, and $\varpi$ is the parallax.}
    \label{tab:basic_info}
\end{table*}

\section{Follow-up} \label{sec:follow-up}

Here, we describe the follow-up spectra that we obtained, the process of measuring metallicities from these spectra, and our constraints on the MS stars' parameters from their spectral energy distributions.  A log of our observations and measured RVs can be found in Appendix \ref{appendix:rvs}. 

\subsection{FEROS}
We obtained 59 spectra with the Fiberfed Extended Range Optical Spectrograph \citep[FEROS;][]{Kaufer1999} on the 2.2\,m ESO/MPG telescope at La Silla Observatory (programs P109.A-9001, P110.A-9014, and P111.A-9003). Some observations used $2\times 2$ binning to reduce readout noise at the expense of spectral resolution; the rest used $1\times 1$ binning. The resulting spectra have resolution $R\approx 40,000$ ($2\times 2$ binning) and $R\approx 50,000$ ($1\times 1$ binning). Exposure times ranged from 1200 to 1800 seconds.

We reduced the data using the CERES pipeline \citep{Brahm2017}, which performs bias-subtraction, flat fielding, wavelength calibration, and optimal extraction. The pipeline measures and corrects for small shifts in the wavelength solution during the course a night via simultaneous observations of a ThAr lamp obtained with a second fiber. We first calculate RVs by cross-correlating a synthetic template spectrum with each order individually and then report the mean RV across 15 orders with wavelengths between 4500 and 6700\,\AA.  We calculate  the uncertainty on this mean RV from the dispersion between orders; i.e., $\sigma_{{\rm RV}}={\rm std}\left({\rm RVs}\right)/\sqrt{15}$. We used a Kurucz spectral template from the \texttt{BOSZ} grid \citep{Bohlin2017} matched to the effective temperature of each star, with log(g) = 4.5 and solar metallicity.

\subsection{TRES}
We obtained 34 spectra using the Tillinghast Reflector Echelle Spectrograph \citep[TRES;][]{Furesz2008} mounted on the 1.5\,m Tillinghast Reflector telescope at the Fred Lawrence Whipple Observatory (FLWO) atop Mount Hopkins, Arizona. TRES is a fibrefed echelle spectrograph with a wavelength range of 390–910 nm and spectral resolution $R\approx 44,000$ ($1\times 1$ binning). Exposure times ranged from 1800 to 3600 seconds. We extracted the spectra as described in \citet{Buchhave2010}.  

As with the FEROS data, we measured RVs by cross-correlating the normalized spectra from each of 31 orders with a Kurucz spectrum template, and we estimate RV uncertainties from the dispersion between RVs measured from different orders; i.e., $\sigma_{{\rm RV}}={\rm std}\left({\rm RVs}\right)/\sqrt{31}$.

\subsection{MIKE}

% - MIKE spectra

We observed J2117+0332 and J2034-5037 with the Magellan Inamori Kyocera Echelle (MIKE) spectrograph on the Magellan 2 telescope at Las Campanas Observatory \citep{Bernstein2003SPIE}. We used the 0.7'' slit with an exposure of 600s. This yielded a spectral resolution $R \sim 35,000$ ($2\times 2$ binning) and typical SNR of $\sim$ 35 and 16 per pixel on the red and blue side respectively. The total wavelength coverage was $\sim 3330 - 9680$ \AA \ (though we only used spectra below 6850 \AA \ to avoid telluric line contamination when computing the metallicity). The spectra were reduced with the MIKE Pipeline using CarPy \citep{Kelson2000ApJ, Kelson2003PASP}. We flux-calibrated the spectra using a standard star and merged the orders into a single spectrum, weighting by inverse variance in the overlap regions. We co-added the two spectra obtained for J2034-5037 across two nights (HJD 2460092.7715 and 2460118.7732) in the same way. J2117+0332 was observed once (HJD 2460118.8313).

\subsection{Metallicities} \label{ssec:met}

Measuring metallicities of the MS stars is important for constraining their masses and ages. 

\subsubsection{SPC}
We fit the TRES spectra using the Stellar Parameter Classification (SPC) tool \citep{Buchhave2012Natur}. This code cross-correlates a grid of synthetic spectra with each observed spectrum in the wavelength range of 5050 to 5360 \AA, centered on the  Mg I b triplet. It then fits the peaks of the cross-correlation function with a three dimensional third order polynomial to return best-fit values of effective temperature $T_{\rm eff}$, surface gravity $\log g$, and metallicity [M/H] that may lie in between the spacings of the grid. 

As described in the supplementary material of \citet{Buchhave2012Natur}, given systematic uncertainties in the synthetic stellar spectra, error floors on the derived [M/H] and $T_{\rm eff}$ values are $\sim$ 0.08 dex and $\sim$ 50 K, respectively \citep[See also ][]{Furlan2018ApJ}. We report these floors in in Table \ref{tab:mets}. 

\subsubsection{BACCHUS}
For the MIKE and FEROS spectra, we used the Brussels Automatic Code
for Characterizing High accUracy Spectra \citep[BACCHUS;][]{Masseron2016ascl, Hayes2022ApJS}. This code performs 1D LTE spectral synthesis to determine stellar parameters. It carries out normalization by linearly fitting the continuum 30\AA \ around a line. It then uses several methods to compare each line of the observed spectrum to that of synthetic spectra to calculate an abundance. 
The effective temperature, surface gravity, and microturbulence are estimated by determining values that result in null trends between the inferred abundances of a given element against the excitation potential, ionization potential, and equivalent widths, respectively. The metallicity [Fe/H] is the mean Fe abundance calculated over lines in the VALD atomic linelist \citep{Piskunov1995A&AS, Ryabchikova2015PhyS} with a wavelength coverage of 4200 to 9200\AA. We assume the detailed abundance pattern traces solar values. The errors reported by BACCHUS represent the scatter in the implied abundances between the different lines and methods of abundance calculations but do not take into account other systematic uncertainties \citep{Hayes2022ApJS}.

\subsubsection{Gaia XP}
We also compare the values measured with SPC and BACCHUS to those calculated by \citet{Andrae2023ApJS} using the {\it Gaia} XP very low-resolution spectra. These authors derive $T_{\rm eff}$, log(g), and [M/H] for 175 million stars with XP spectra published in  DR3. Although the spectra from which these parameters are derived have low resolution, \citet{Andrae2023ApJS} demonstrated that their reported metallicities are accurate to within better than 0.1 dex for bright and nearby stars like our targets with temperatures within the range of our sample.

\subsubsection{Results}

The metallicities and effective temperatures obtained from  spectral fitting are summarized in Table \ref{tab:mets}. The metallicities range from $-0.15$ to 0.20 dex. For J2117+0332, we see that the metallicities from SPC and BACCHUS are in agreement. The {\it Gaia} XP metallicites are not used in our analysis in the following sections but provide a useful comparison point. Most of our [M/H] measurements are consistent with the {\it Gaia} XP measurements from  \citet{Andrae2023ApJS} within $ 1\sigma$. The good agreement between the three metallicities shows that XP metallicities are likely sufficiently accurate for analysis of larger samples in cases where high resolution follow-up would be prohibitively expensive. We also add a column for the best-fit [Fe/H] values from our SED fitting (Section \ref{ssec:sed}), which uses the SPC and BACCHUS metallicities as a prior.

\begin{table*}
    \makebox[\textwidth][c]{
        \begin{tabular}{c|cccc|cccc}
            \hline
              & \multicolumn{4}{c}{[Fe/H]} & \multicolumn{4}{c}{$T_{\rm{eff}}$ [K]} \\
            \cmidrule(lr){2-5}\cmidrule(lr){6-9}
            name & SPC & BACCHUS  & Gaia XP & SED  & SPC & BACCHUS  & Gaia XP & SED \\ \hline
            J2117+0332 & -0.24 $\pm$ 0.08 & -0.284 $\pm$ 0.18  & -0.380 & -0.22 $\pm$ 0.06 & 6029 $\pm$ 50 & 6152 +/- 79 & 6111.0 & 6226 $\pm$ 19 \\
            J1111+5515  & -0.15  $\pm$ 0.08 & - & -0.172 & -0.17 $\pm$ 0.06 & 5987 $\pm$ 50 & - & 6006.3 &  6190 $\pm$ 22 \\
            J1314+3818  & -0.39  $\pm$ 0.08 & -  &-0.291 & -0.34 $\pm$ 0.05  & 4707 $\pm$ 50 & - & 4700.2 & 4684 $\pm$ 13 \\
            J2034-5037  & - & -0.346  $\pm$ 0.078 &-0.352 & -0.19 $\pm$ 0.06 & - & 5789 $\pm$ 17 & 5758.8  & 5856 $\pm$ 20 \\
            J0107-2827  & - & 0.198  $\pm$ 0.127  & 0.244 & 0.04 $\pm$ 0.07 & -  & 5524 $\pm$ 51 & 5330.4 & 5387 $\pm$ 21 \\
            \hline
        \end{tabular}
    }
    \caption{Comparison of the metallicities and $T_{\rm{eff}}$ obtained from various methods.}
    \label{tab:mets}%
\end{table*}%

\subsection{Light curves} \label{ssec:lightcurves}

We retrieved observed light curves for our objects from the All-Sky Automated Survey for Supernovae \citep[ASAS-SN; ][]{Shappee2014ApJ, Kochanek2017PASP}. We used the $V$ band data, for which the number of photometric points ranged from 1609 to 3194 across the five objects. The typical uncertainty in normalized flux is $\sim 0.01$. To search for periodic variability, we computed Lomb-Scargle periodograms of these light curves \citep{Lomb1976Ap&SS, Scargle1982ApJ, AstropyCollaboration2022ApJ}. We did not find any significant periodicities beyond the lunar cycle and sidereal day. This allows us to rule out periodic variability with amplitude greater than the strongest noise peaks, which have amplitude $\sim 0.002 - 0.003$ for all objects except J1314+3813, where they have amplitude $\sim 0.006$.

\subsection{SED fitting} \label{ssec:sed}

We constructed broadband spectral energy distributions (SEDs) of our targets using synthetic {\it ugriz} SDSS photometry calculated from {\it Gaia} XP spectra \citep{GaiaCollaboration2022arXiv} (with the exception of J2117+0332 where actual SDSS photometry was available and used instead; \citealt{Padmanabhan2008ApJ}), 2MASS {\it JHK} photometry \citep{Skrutskie2006AJ}, and {\it WISE} $W_1W_2W_3$ photometry \citep{Wright2010AJ}. We obtained $E(B-V)$ for each object using the \citet{Lallement2022A&A} 3D dust map for declinations below -30$^{\circ}$ and the Bayestar2019 3D dust map \citep{Green2019ApJ} for declinations above -30$^{\circ}$. These are given in Table \ref{tab:sed_params}. We assume a \citet{Cardelli1989ApJ} extinction law with $R_V = 3.1$. The Bayestar2019 map provides $E(g-r)$ which is approximately equal to $E(B-V)$ \citep{Schlafly2011ApJ}, while the \citet{Lallement2022A&A} map provides the extinction $A_0$ at 550 nm which we take to be $A_V$. As all our objects are relatively nearby with $E(B-V) < 0.05$, the uncertainties in these extinction values do not dominate the uncertainties in the final fitted parameters. 

We do not attempt to account for flux contributions from the WD companions, which must be very small (given their high masses) and faint in the optical. We justify this assumption in Appendix \ref{appendix:wd_sed}, where we show that even very hot WDs with $T_{\rm eff} = 60,000$ K would not significantly contribute to the photometry of all but one of our targets. For the one exception, J1314+3818, we find that a WD with  $T_{\rm eff} \gtrsim$ 30,000 K could contribute to the $u$-band photometry, so we conservatively excluded the $u-$band measurement from our fit.

We fit the SEDs using \texttt{MINEsweeper}  \citep{Cargile2020ApJ}, a code designed for joint modeling of stellar photometry and spectra. We only use the code's photometric modeling capabilities but place a prior on the present-day surface metallicity from spectroscopy. The free parameters to be fit are each star's parallax, mass $M_{\star}$, initial metallicity [Fe/H]$_{\rm init}$, and Equivalent Evolutionary Phase (EEP, a monotonic function of age; see \citealt{Dotter2016ApJS}).  From each set of parameters, \texttt{MINEsweeper} generates a predicted SED and photometry in specified filters using neural network interpolation. We use \texttt{emcee}, a Python  Markov chain Monte Carlo (MCMC) sampler \citep{Foreman-Mackey2013PASP}, to sample from posterior. Constraints from fitting each source's SED are listed in Table~\ref{tab:sed_params}.

We note that \texttt{MINEsweeper} constrains the \textit{initial} metallicity, which is not identical to the present-day surface value measured from spectroscopy. For our targets, the difference between initial and present-day surface metallicity is a result of atomic diffusion, where heavier elements settle out of the atmosphere over time \citep{Dotter2017ApJ}. The present-day surface metallicity [Fe/H] is predicted by the isochrones given a set of $M_{\star}$, [Fe/H]$_{\rm init}$, and EEP, so the spectroscopic metallicities found in Section \ref{ssec:met} are used to add a Gaussian constraint on [Fe/H] to the likelihood. While values for $T_{\rm eff}$ are also obtained from spectral analysis (Table \ref{tab:mets}), given the degeneracy that can exist between $T_{\rm eff}$ and $\log g$ in spectroscopic fits and the high quality of the SED fits, we do not use them to constrain the outputs here.

Putting everything together, the final likelihood function is:
\begin{equation} 
\begin{split}
\ln L = & -\frac{1}{2} \sum_i \frac{\left(\rm mag_{pred, i} - mag_{obs, i}\right)^2} {\sigma^2_{\rm mag, i}} 
 -\frac{1}{2} \frac{\left(\rm [Fe/H]_{pred} - [Fe/H]_{obs}\right)^2}{\sigma^2_{\rm [Fe/H]}} \\
 & -\frac{1}{2} \frac{\left(\rm \varpi_{pred} - \varpi_{obs}\right)^2}{\sigma^2_{\varpi}},
\end{split}
\label{eqn:lnL_sed}
\end{equation} 
where ``mag" stands for apparent magnitudes and the summation is over the appropriate photometric filters for each object, $\sigma_{X}$ is the error on the observed value of some quantity $X$, and [Fe/H] is the present-day surface metallicity. We set a floor on $\sigma_{\rm mag}$ of 0.02 dex (given possible calibration issues) to avoid underestimating the errors.

We report the medians of the marginalized posterior distributions for each parameter in Table \ref{tab:sed_params}. $M_{\star}$, EEP, and  [Fe/H]$_{\rm init}$ are the parameters directly fitted by \texttt{MINEsweeper}, while [Fe/H], $T_{\rm eff}$, and $R$ are calculated from the isochrones corresponding to the fitted parameters. The fit to parallax and the reported errors are described in Section \ref{sssec:parallax_errs}. We also list constraints on [Fe/H] and $T_{\rm eff}$ for comparison with the values measured from spectroscopy  (Table \ref{tab:mets}).

\begin{figure*}
    \centering
    \includegraphics[width=0.95\textwidth]{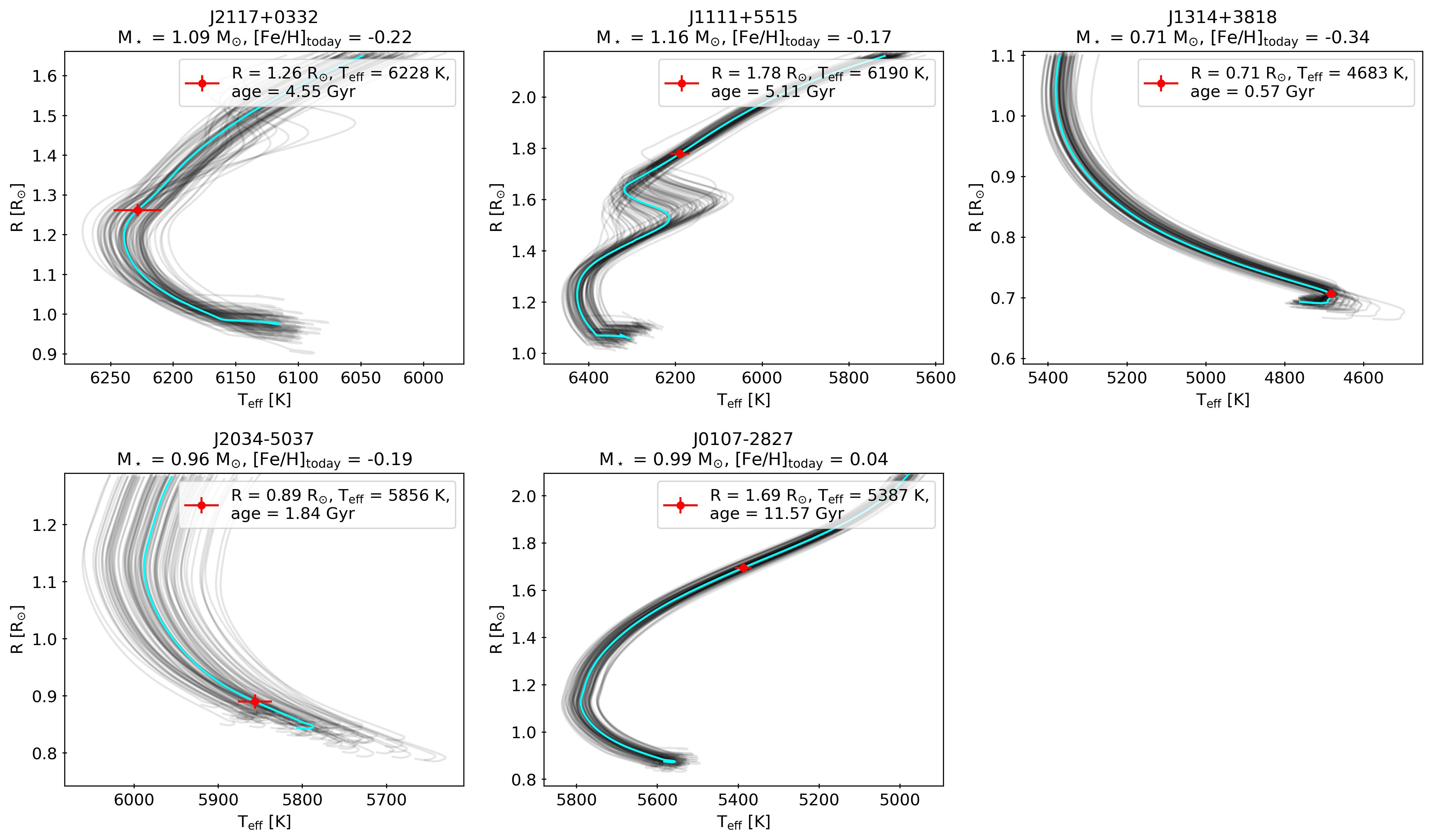}
    \caption{MIST isochrones for all of our objects. In each plot, gray lines are the isochrones given the masses and metallicities of 100 randomly chosen posterior samples from the photometric fitting, while the cyan line is that of the best-fit parameters. The red point marks the present location of the object and the error bars are the standard deviation in the radii and temperatures of the posteriors at the corresponding ages.}
    \label{fig:isochrones}
\end{figure*}

Figure \ref{fig:isochrones} shows MIST isochrones corresponding to the stellar parameters of 100 random posterior samples (gray). The cyan lines show the best-fit parameters and the red point marks the present inferred parameters of the MS stars. The labels also indicate the stellar ages, which range from 1.84 to 11.57 Gyrs. %We see that the error bars are often smaller than the visual spread of the isochrones (especially for the radius), because the age (i.e. EEP) changes accordingly to give $T_{\rm eff}$ and $R$ which fit the observed photometry. 
Two systems, J111+5515 and J0107-2827, host stars that have slightly evolved off the MS. This is likely to be the result of selection bias, as evolved stars are brighter and thus over-represented in magnitude-limited samples. In addition, we assumed that stars were on the MS when estimating their masses in our initial selection of targets for follow-up.  These initial estimates were moderately overestimated for evolved stars, leading to overestimated companion masses. Since we targeted massive companions -- and massive companions are intrinsically rare -- we expect evolved MS stars to be preferentially selected.

The observed and predicted SEDs are shown in Figure \ref{fig:seds}. The model SEDs plotted were generated using \texttt{pytstelllibs}\footnote{https://mfouesneau.github.io/pystellibs/} with the best-fit parameters as inputs. We have checked that these models give roughly consistent photometry to that predicted with MINEsweeper which does not itself return a continuous SED. The residuals of the photometry typically lie within 0.1 mag. As mentioned above, for J1314+3818, we found that WDs with $T_{\rm eff} \gtrsim 30000 $ K would significantly contribute to the SDSS $u$-band photometry (Appendix \ref{appendix:wd_sed}), so we excluded this point in fitting. GALEX NUV observations are shown in Figure \ref{fig:seds}, but these were also excluded from the fitting for all targets to avoid potential contamination from the WD companion. We investigate the expected contributions of  WD companions to the NUV photometry in Appendix \ref{appendix:wd_sed}. There we show that the observed NUV photometry is consistent with no contributions from the WDs for all our targets. This places an upper limit on the effective temperatures of the WDs, ranging from from $T_{\rm eff,\,WD} < 7750\,{\rm K}$ for J1314+3818  to $T_{\rm eff,\,WD} < 60,000\,{\rm K}$ for J1111+5515.

\subsubsection{A wide tertiary}

\begin{figure*}
    \centering
    \includegraphics[width=0.95\textwidth]{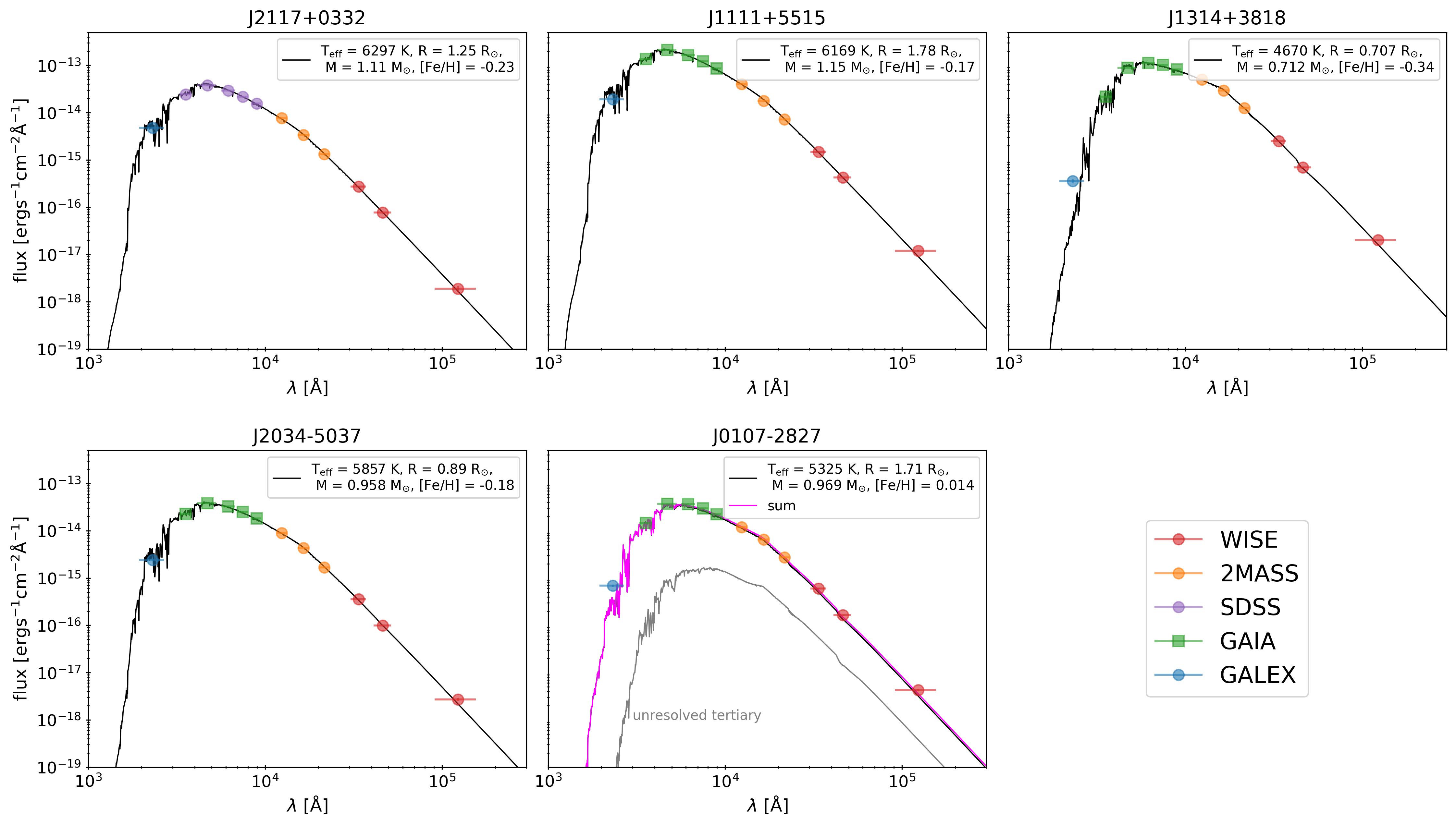}
    \caption{Model SEDs for all five objects using the best-fit parameters from fitting the photometry using MINEsweeper, compared with the observations. J0107-2827 has a visible third companion whose SED is shown in gray. The magenta line is the sum of the two luminous stars (the black line corresponding to the more luminous star lies close to, but under, the magenta line). Note that while GALEX points are plotted, these were not used in the photometric fitting to avoid possible contamination from the WD companions.}
    \label{fig:seds}
\end{figure*}

One system, J0107-2827, has a resolved tertiary separated by a distance of 2.21 arcseconds, corresponding to a projected physical separation of 1033 AU  (Gaia DR3 ID 5033197892724532608). The consistency in the parallaxes and proper motions of the two sources  make it highly likely that they are in fact physically bound, as opposed to a chance alignment \citep[e.g.][]{El-Badry2021MNRAS}.  While the source is resolved by {\it Gaia} in the $G$ band, the XP, 2MASS, and {\it WISE} photometry of the two sources are likely all unresolved, so we model its SED as a sum of two luminous stars. Assuming the tertiary is on the main sequence, its $G-$band absolute magnitude of $\sim 7.3$ (calculated using the reported apparent magnitude of 15.7) corresponds to a mass of approximately $0.67\,M_{\odot}$. We  assume solar metallicity (consistent with the initial metallicity we infer for the primary) and an age of $\sim 6$ Gyr. Using these parameters, we generated photometry for this third star (gray line in Figure \ref{fig:seds}) which we added to the model primary (black). This sum (magenta) was fit to the observations.

\subsubsection{Effect of potentially underestimated parallax errors} \label{sssec:parallax_errs}

Our fitting also leaves the parallax, $\varpi$, free, allowing us to propagate parallaxes uncertainties through to the stellar parameters. From Table \ref{tab:basic_info}, we see that the Renormalised Unit Weight Errors (RUWEs) from {\it Gaia} DR3 for several of our objects are above 1.4, which may indicate that the reported parallax uncertainties are underestimated \citep{Lindegren2018GAIA} as a result of orbital motion, which is not accounted for in the {\it Gaia} single-star astrometric model.

To estimate more realistic parallax uncertainties, we carry out the following analysis. We select  sources from the {\it Gaia} NSS catalog with Orbital or AstroSpectroSB1 solutions. In addition to the ``single-star model'' parallaxes reported for these solutions in the \texttt{gaia\_source} table, these sources have improved parallaxes from astrometric solutions that account for wobble induced by their binarity \citep{GaiaCollaboration2023A&A, Halbwachs2023A&A}. From these, we select those with phot\_g\_mean\_mag < 13 and RUWE values comparable to our targets. We then calculate the standard deviation of the difference between the parallaxes reported from single-star solutions (gaia\_source) and binary solution (NSS catalog). We found a standard deviation of 0.104 mas for 13,132 sources with RUWE between 1.4 and 2, and a standard deviation of 0.181 mas for 18,351 sources with RUWE between 2 and 3. The maximum RUWE for our objects is 2.94 (Table \ref{tab:basic_info}). 

Based on these values, we re-run the SED fitting with increased parallax uncertainties of 0.2 mas for J2034-5037 (with a RUWE > 2) and 0.1 mas for the remaining four objects. We find no significant changes to the best-fit values of the parameters but a general increase in the uncertainties (i.e. the standard derivations of the parameters from the posterior). We report these inflated uncertainties in Table \ref{tab:sed_params}. Note that in Section \ref{ssec:wd_masses}, we also set an uncertainity floor of $\pm 0.05\,M_{\odot}$ on $M_{\star}$ to obtain conservative errors on the inferred masses of the WDs. 

% Table of solutions - Minit, Metallicity, Radius etc. 
\begin{table*} 
     \centering
     \begin{tabular}{c c c c c c c c c c}
          \hline
          Name & $E(B-V)$ & $M_{\star}$ [$M_{\odot}$] & EEP &  $\varpi$ [mas] & d [pc] & [Fe/H]$_{\rm init}$ &  [Fe/H] & $T_{\rm eff}$ [K] & $R$ [$R_{\odot}$] \\
          \hline
         J2117+0332 & 0.045 & 1.11 $\pm$ 0.03 & 383.26 $\pm$ 16.40 & 1.96 $\pm$ 0.10 & 511 $\pm$ 25 & -0.09 $\pm$ 0.05 & -0.23 $\pm$ 0.05 & 6297 $\pm$ 20 & 1.25 $\pm$ 0.06 \\
         J1111+5515 & 0.009 & 1.15 $\pm$ 0.02 & 444.90 $\pm$ 2.92 & 3.24 $\pm$ 0.10 & 309 $\pm$ 9 & -0.09 $\pm$ 0.06 & -0.17 $\pm$ 0.06 & 6169 $\pm$ 22 & 1.78 $\pm$ 0.06 \\
         J1314+3818 & 0.000 & 0.71 $\pm$ 0.01 & 276.03 $\pm$ 35.93 & 12.45 $\pm$ 0.10 & 80 $\pm$ 1 & -0.33 $\pm$ 0.04 & -0.34 $\pm$ 0.05 & 4670 $\pm$ 13 & 0.71 $\pm$ 0.01 \\
         J2034-5037 & 0.024 & 0.96 $\pm$ 0.02 & 321.80 $\pm$ 52.27 & 3.24 $\pm$ 0.17 & 308 $\pm$ 16 & -0.16 $\pm$ 0.06 & -0.18 $\pm$ 0.06 & 5857 $\pm$ 19 & 0.89 $\pm$ 0.05 \\
         J0107-2827 & 0.027 & 0.97 $\pm$ 0.03 & 459.26 $\pm$ 1.39 & 2.14 $\pm$ 0.08 & 468 $\pm$ 17 & 0.08 $\pm$ 0.07 & 0.01 $\pm$ 0.06 & 5325 $\pm$ 19 & 1.71 $\pm$ 0.07 \\
          \hline  
     \end{tabular}
     \caption{Best-fit parameters from SED fitting. We have also added the extinction $E(B-V)$ for all objects -- these were taken from 3D dust maps and  have uncertainties of $\sim 0.02$ mag. $M_{\star}$ is the mass of the luminous star, EEP is the Equivalent Evolutionary Phase (related to its age), [Fe/H]$_{\rm init}$ is the initial metallicity of the star, and $\varpi$ is the parallax with d being the corresponding distance in pc. For $M_{\star}$, we set an uncertainity floor of $\pm 0.05\,M_{\odot}$ when calculating WD masses in Section \ref{ssec:wd_masses}. The other parameters are inferred from the isochrone corresponding to the fitted parameters, where [Fe/H] is the present-day metallicity, $T_{\rm eff}$ is the effective temperature, and $R$ is the radius.}
     \label{tab:sed_params}
 \end{table*}
         
\section{Orbital fits and constraints on the unseen companion} \label{sec:rvs}

We fit the FEROS and TRES RVs with a Keplerian model using \texttt{emcee}. The free parameters of the fit are the orbital period $P_{\rm orb}$, periastron time $T_p$, eccentricity $e$, RV semi-amplitude $K_{\star}$, argument of periastron $\omega$, and center-of-mass RV $\gamma$. In the case of J2117+0332, where we have RVs from two different instruments, we also fit for an RV offset between the two instruments as an additional parameter. We set broad, uniform priors on all parameters. The likelihood function is defined as:
\begin{equation} \label{eqn:logL}
    \ln L = - \frac{1}{2}  \sum_i \frac{\left({\rm RV}_{\rm pred}\left(t_i\right) - {\rm RV}_i\right)^2}{\sigma^2_{{\rm RV}, i}}
\end{equation}
where ${\rm RV}_{\rm pred}\left(t_i\right)$ and ${\rm RV}_i$ are the predicted and measured RVs at times $t_i$, and $\sigma^2_{{ \rm RV}, i}$ are the errors in the measurements. 

\begin{figure*}
     \centering
     \includegraphics[width=0.33\textwidth]{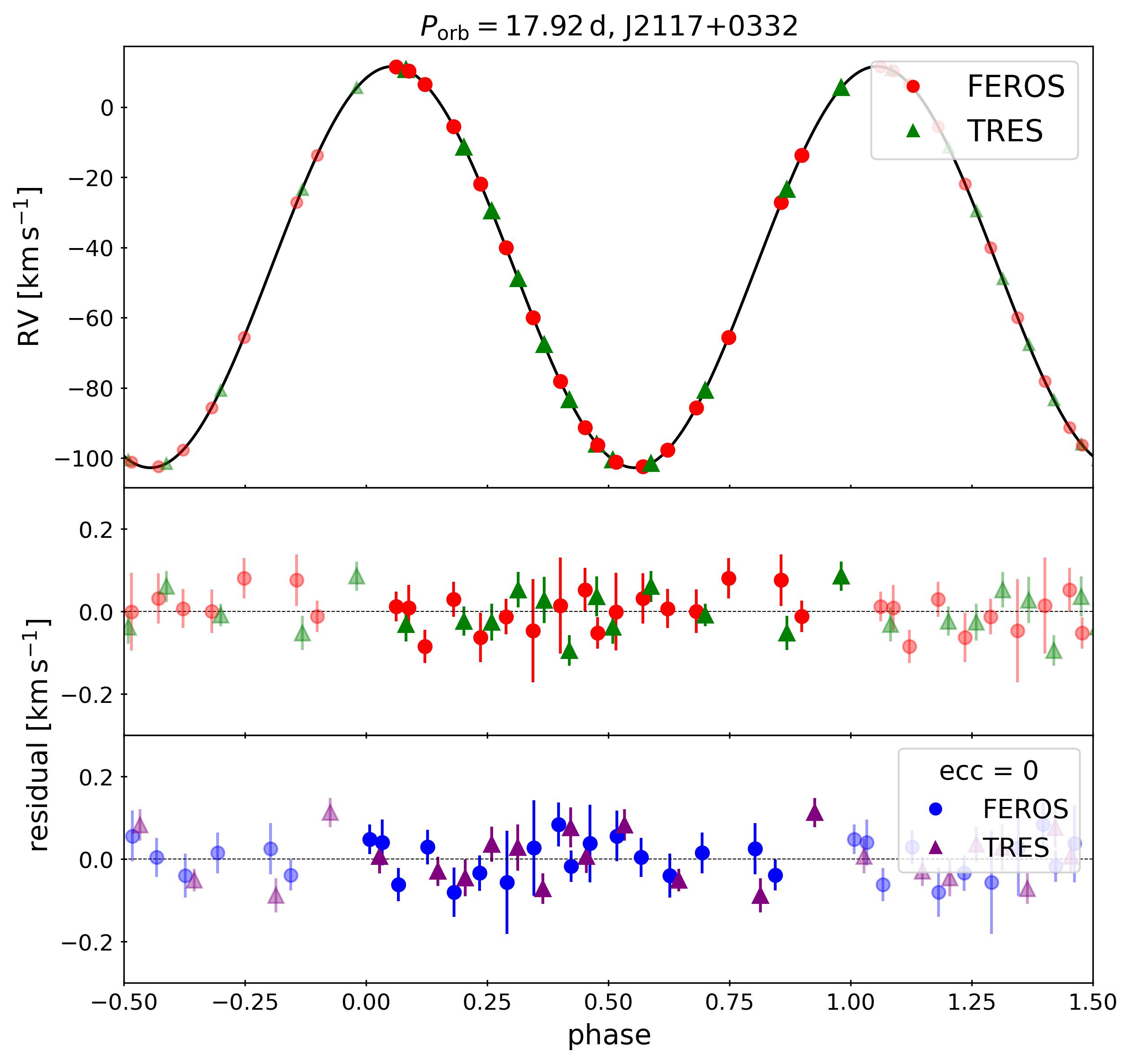}
     \includegraphics[width=0.33\textwidth]{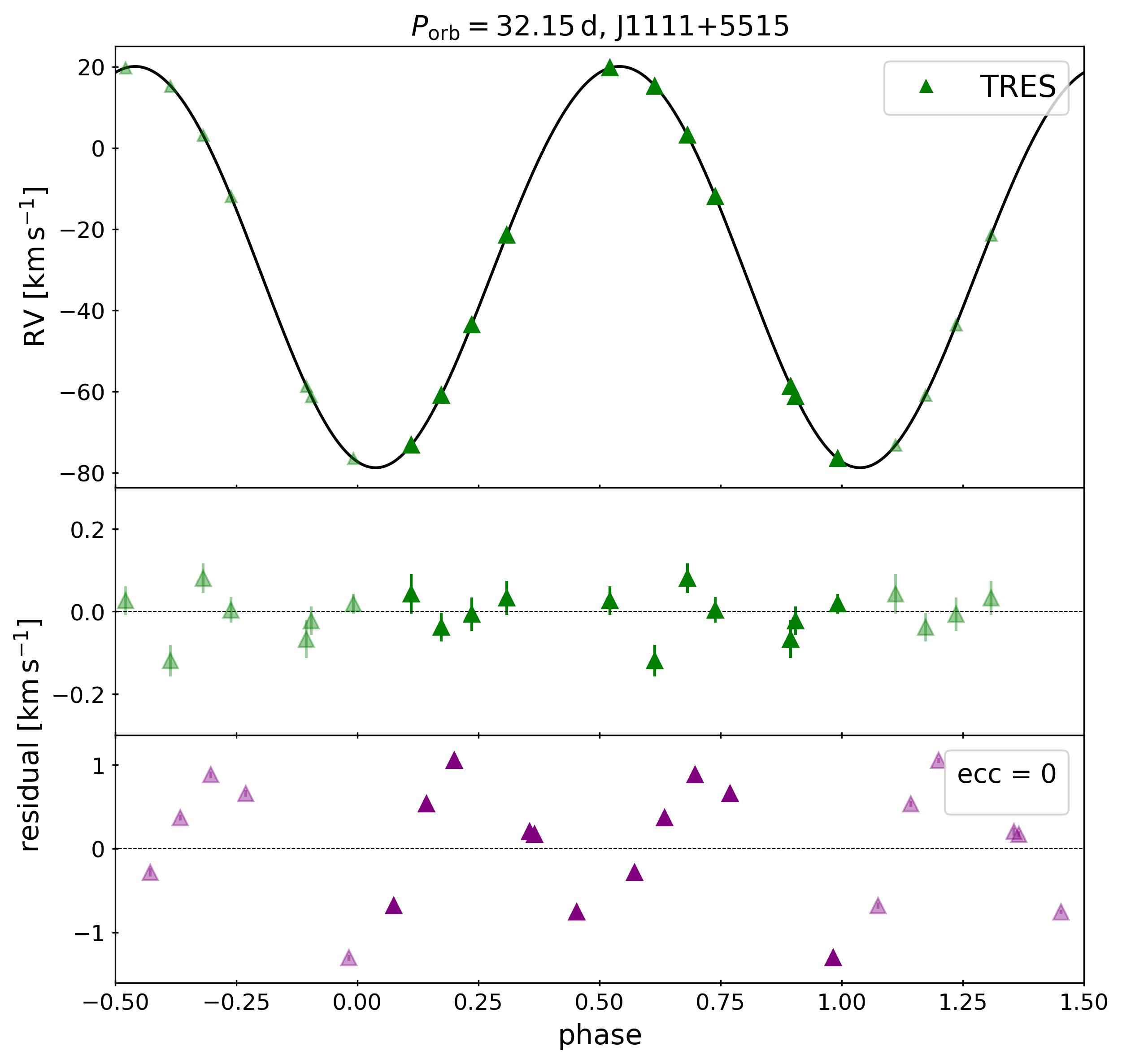} 
     \includegraphics[width=0.33\textwidth]{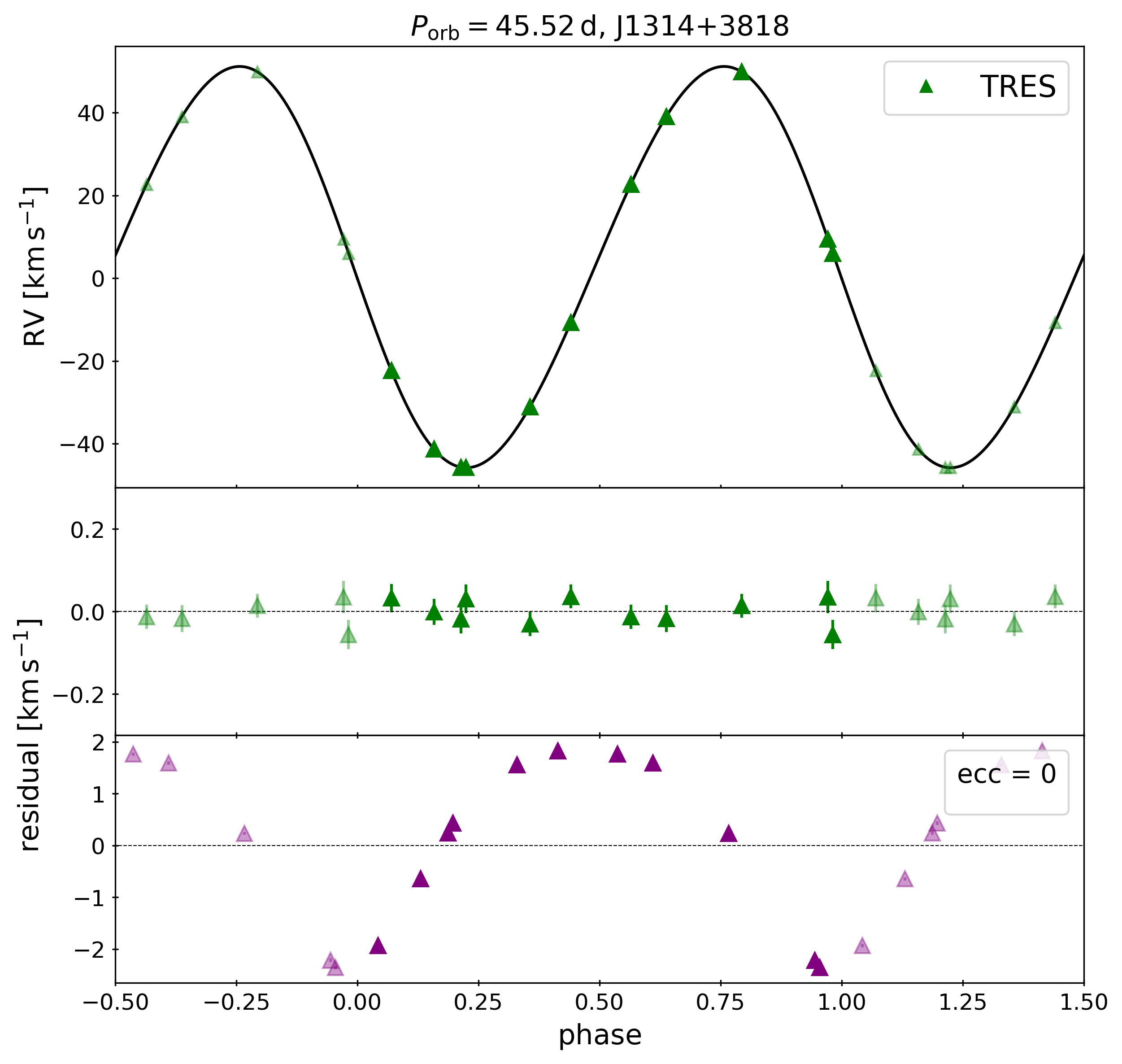} \\
     \includegraphics[width=0.33\textwidth]{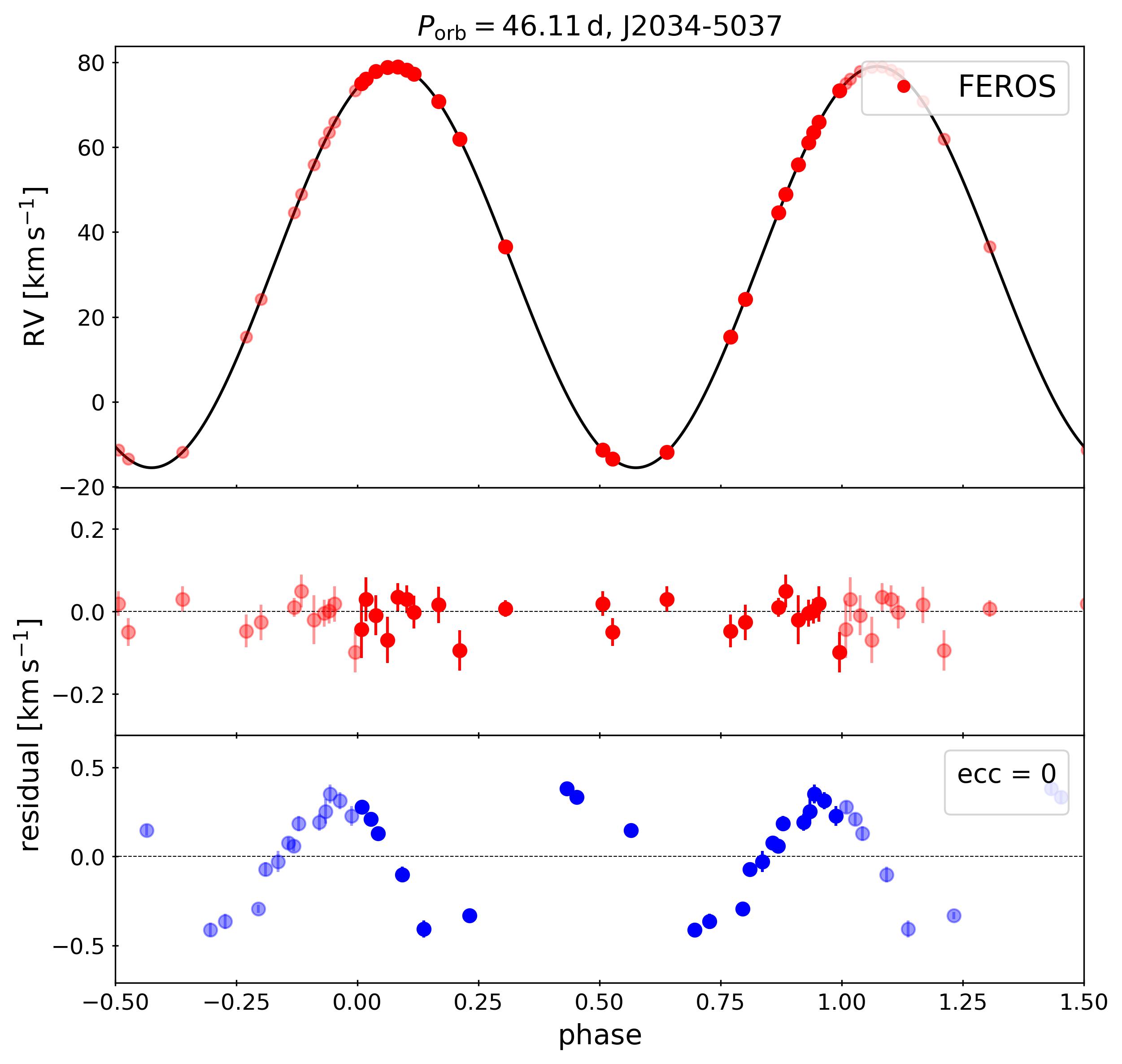}
     \includegraphics[width=0.33\textwidth]{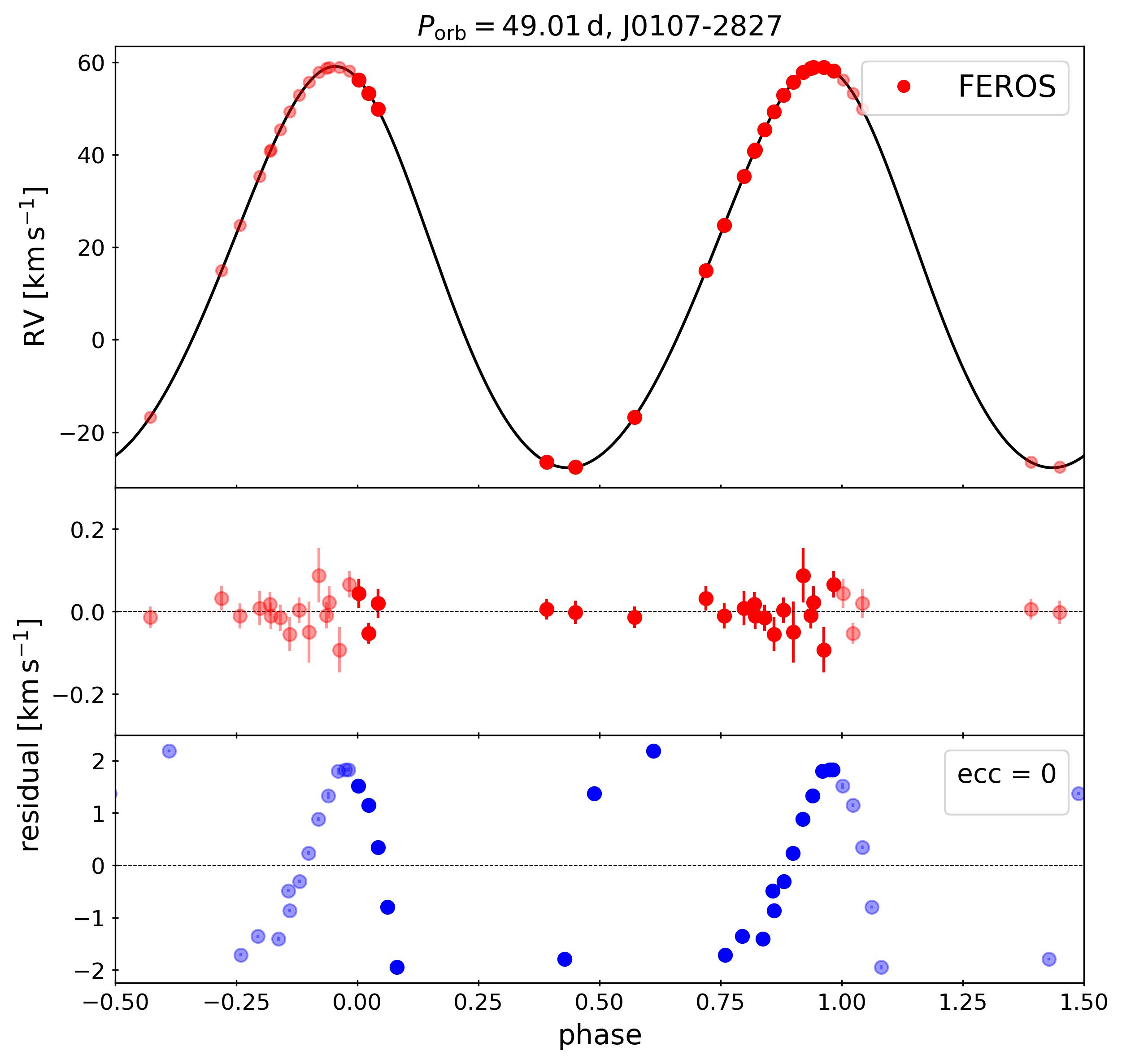}
     \caption{Results of the RV fitting. For each object, the top panel shows the best-fit RV curve over the observed points, the second panel shows the residuals of this fit, and the third panel shows the residuals of the fit with eccentricity set to 0. For all objects (except possibly J2117+0332), the residuals for the model which fits for the eccentricity are significantly smaller than those with eccentricity fixed to zero.}
     \label{fig:RV_curves}
\end{figure*}

The best-fit RV curve for each object is shown in Figure \ref{fig:RV_curves}. Best-fit values for $P_{\rm orb}$, $e$, and $K_{\star}$, along with the implied mass functions $f_m$ given these parameters, are reported in Table \ref{tab:rv_fitting}. For comparison, we also list  the mass functions calculated using the same parameters from the {\it Gaia} DR3 SB1 solutions.

We find that all of the systems have a small but non-zero eccentricity. To confirm that these are significant, we also fit the RVs using a model with eccentricity and $\omega$ fixed to zero. The residuals from the two models are plotted on the second and third panels for each object in Figure \ref{fig:RV_curves}. We see that the residuals from the model with $e = 0$ are obviously larger than those from the model that fits for $e$, with the possible exception of J2117+0332 (which has $e=0.0007\pm 0.0002$) where the difference is more subtle. 

Since eccentricity cannot be negative, observational eccentricities will result in a positive eccentricity bias for orbits that are nearly circular \citep[e.g.][]{Hara2019}. 
For J2117+0332, we generate simulated RVs with the orbital parameters of the $e = 0$ fit at the JDs of our observations, adding to them Gaussian noise with a standard deviation of $0.05\,\rm km\,s^{-1}$. We then fit these RVs with a Keplerian model, which yields $e \sim 0.0003 \pm 0.0002$. This is comparable to the uncertainty on $e$ we find with the measured RVs, and smaller than the measured eccentricity. This experiment provides additional support that the non-zero eccentricity measured for J2117+0332 is real.

% Table of orbital solutions
\begin{table*} 
     \centering
     \begin{tabular}{|c c c c c c|}
          \hline
          Name & $P_{\rm orb}$ [d] & $e$ &  $K_{\star}$ [km/s] &  $f_m$ [$M_{\odot}$] &  $f_{m, G}$ [$M_{\odot}$] \\
          \hline
         J2117+0332 & 17.9239 $\pm$ 0.0001 & 0.0007 $\pm$ 0.0002 & 57.215 $\pm$ 0.011 & 0.3478 $\pm$ 0.0002 & 0.4143 $\pm$ 0.0398 \\
         J1111+5515 & 32.1494 $\pm$ 0.0022 & 0.0217 $\pm$ 0.0003 & 49.435 $\pm$ 0.019 & 0.4021 $\pm$ 0.0004 & 0.3981 $\pm$ 0.0052 \\
         J1314+3818 & 45.5150 $\pm$ 0.0047 & 0.0503 $\pm$ 0.0003 & 48.468 $\pm$ 0.015 & 0.5349 $\pm$ 0.0005 & -- \\
         J2034-5037 & 46.1147 $\pm$ 0.0006 & 0.0079 $\pm$ 0.0002 & 47.299 $\pm$ 0.012 & 0.5056 $\pm$ 0.0004 & 0.6392 $\pm$ 0.0944 \\
         J0107-2827 & 49.0063 $\pm$ 0.0008 & 0.0901 $\pm$ 0.0005 & 43.370 $\pm$ 0.011 & 0.4092 $\pm$ 0.0003 & 0.4175 $\pm$ 0.0275 \\
          \hline  
     \end{tabular}
     \caption{Best-fit orbital parameters from the RV fitting. $f_m$ is the mass function given the other three parameters and $f_{m, G}$ is the mass function given the same parameters from the {\it Gaia} solution. }
     \label{tab:rv_fitting}
 \end{table*}

\subsection{Masses of the unseen companion} \label{ssec:wd_masses}

From the parameters of the RV fitting, we can calculate the mass function, $f_m$, which provides a constraint on the mass of the unseen companion $M_{\rm WD}$. (Note that this notation implies that the unseen companion is a WD which we argue is the most likely scenario in Section~\ref{ssec:alternatives}. We keep this notation here for consistency throughout the paper):
\begin{equation} \label{eqn:massfunc}
    f_m = \frac{M_{\rm WD}^3 \sin^3 i}{\left( M_{\star} + M_{\rm WD} \right)^2} = \frac{P_{\rm orb} K_{\star}^3}{2 \pi G} \left(1 - e^2 \right)^{3/2}
\end{equation}
where $M_{\star}$ is the mass of the luminous star, which we constrained by fitting the SED (Section \ref{ssec:sed}). With just the RVs, the inclination $i$ is not constrained, meaning that for most of our objects, we can only place a lower limit on the WD mass which occurs when $i = 90^{\circ}$ (i.e. when the orbit is ``edge-on'' to our line of sight). 

\begin{figure*}
     \centering
     \includegraphics[width=0.33\textwidth]{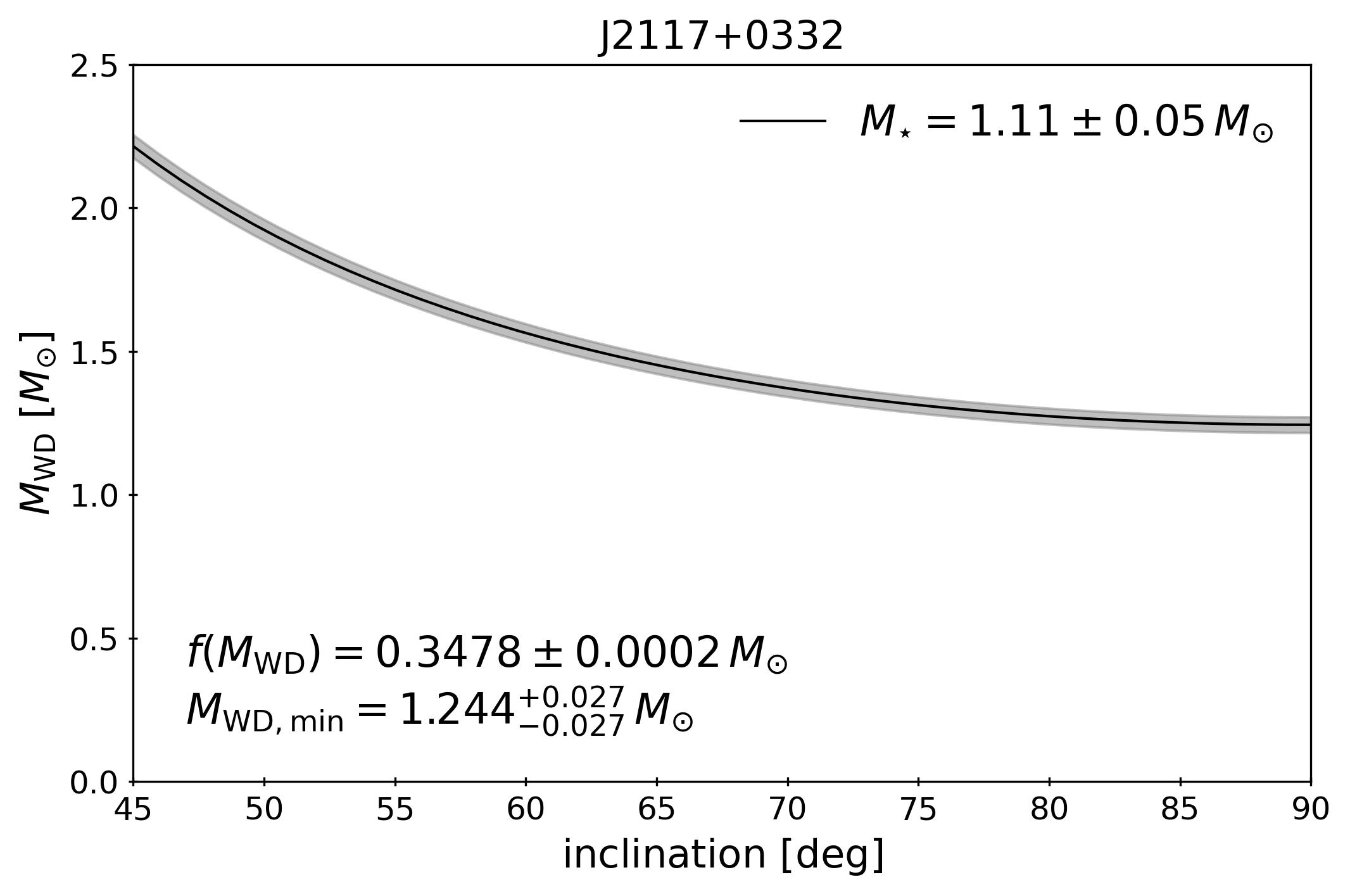}
     \includegraphics[width=0.33\textwidth]{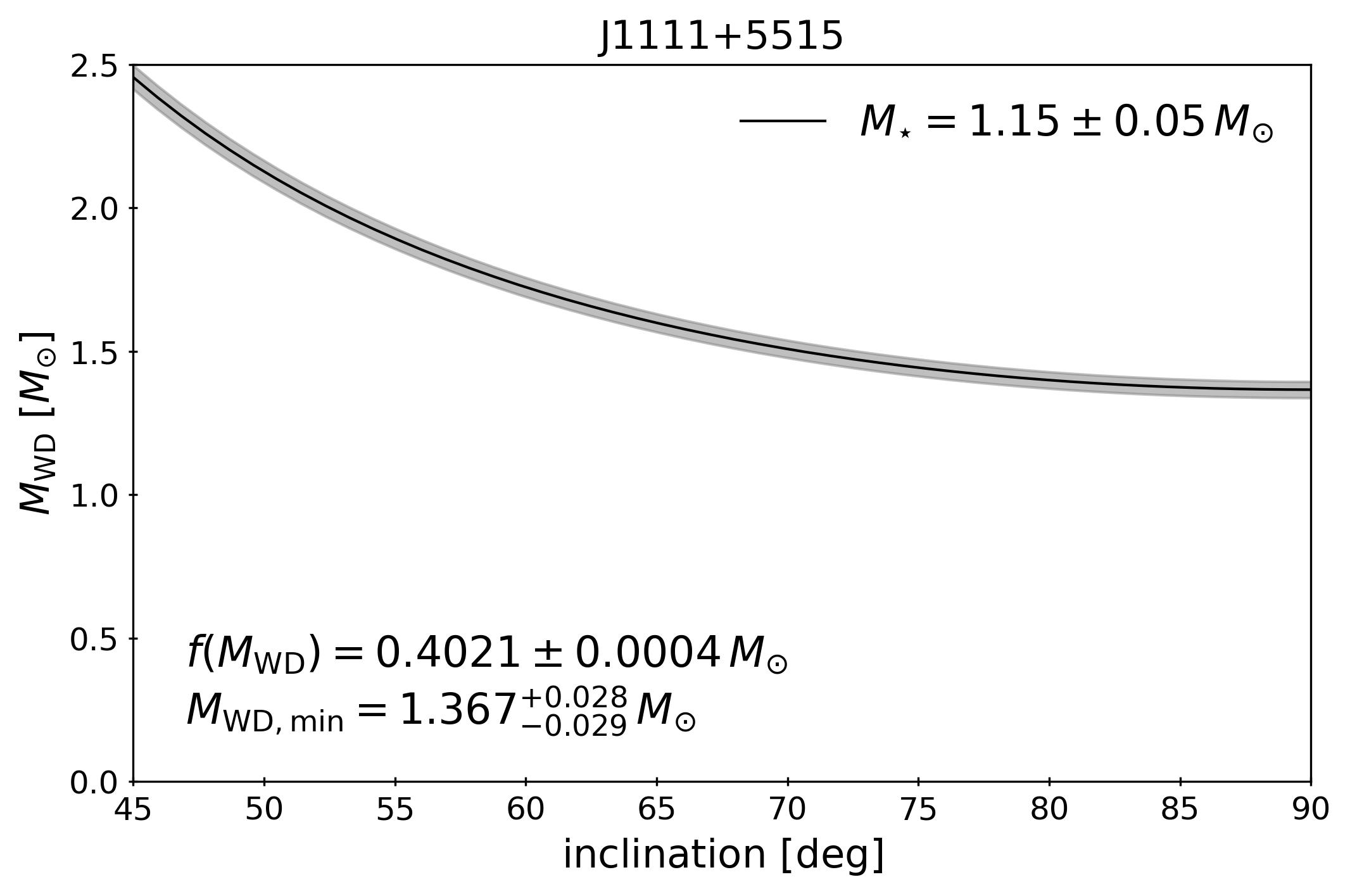} 
     \includegraphics[width=0.33\textwidth]{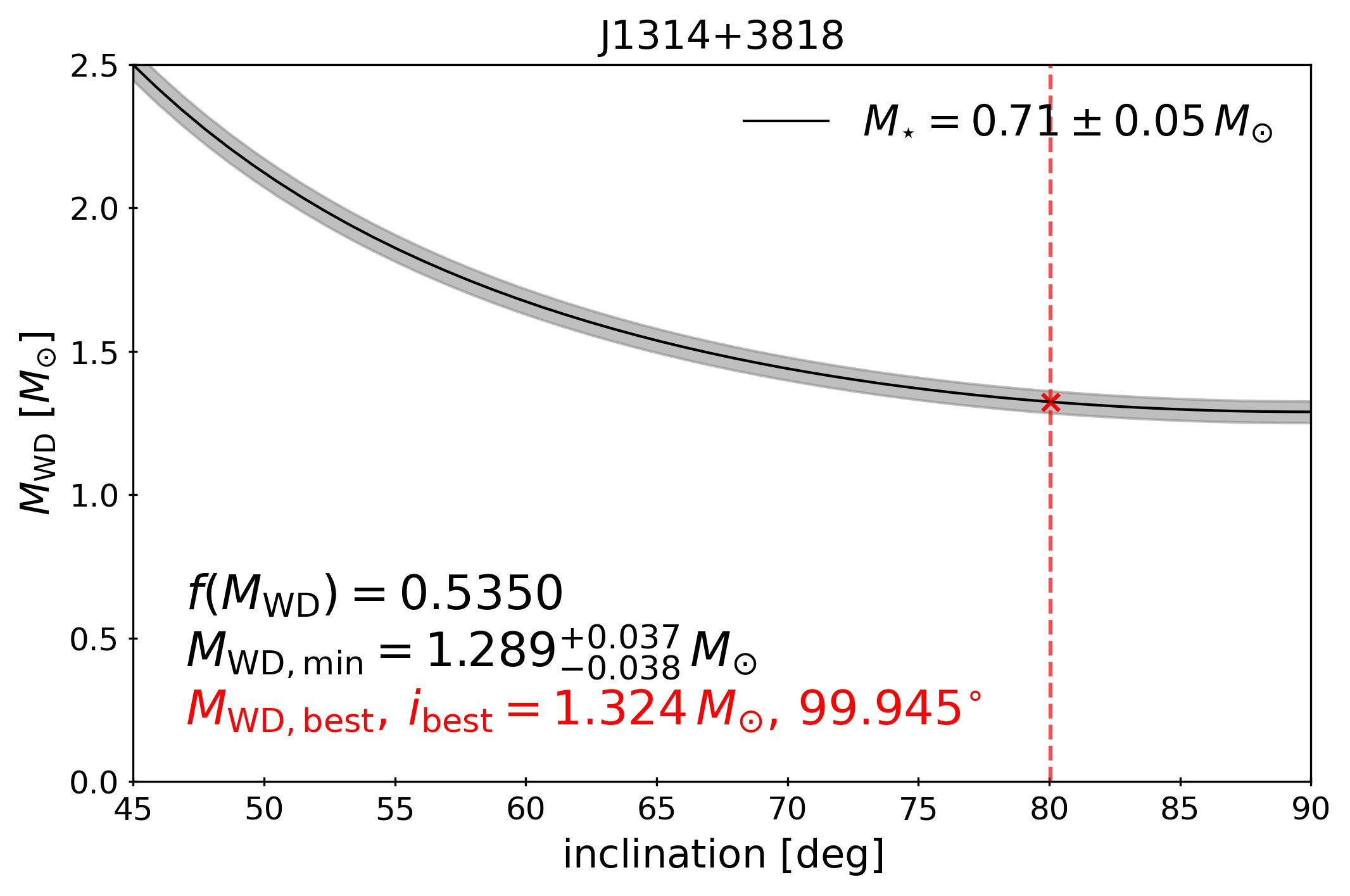} \\
     \includegraphics[width=0.33\textwidth]{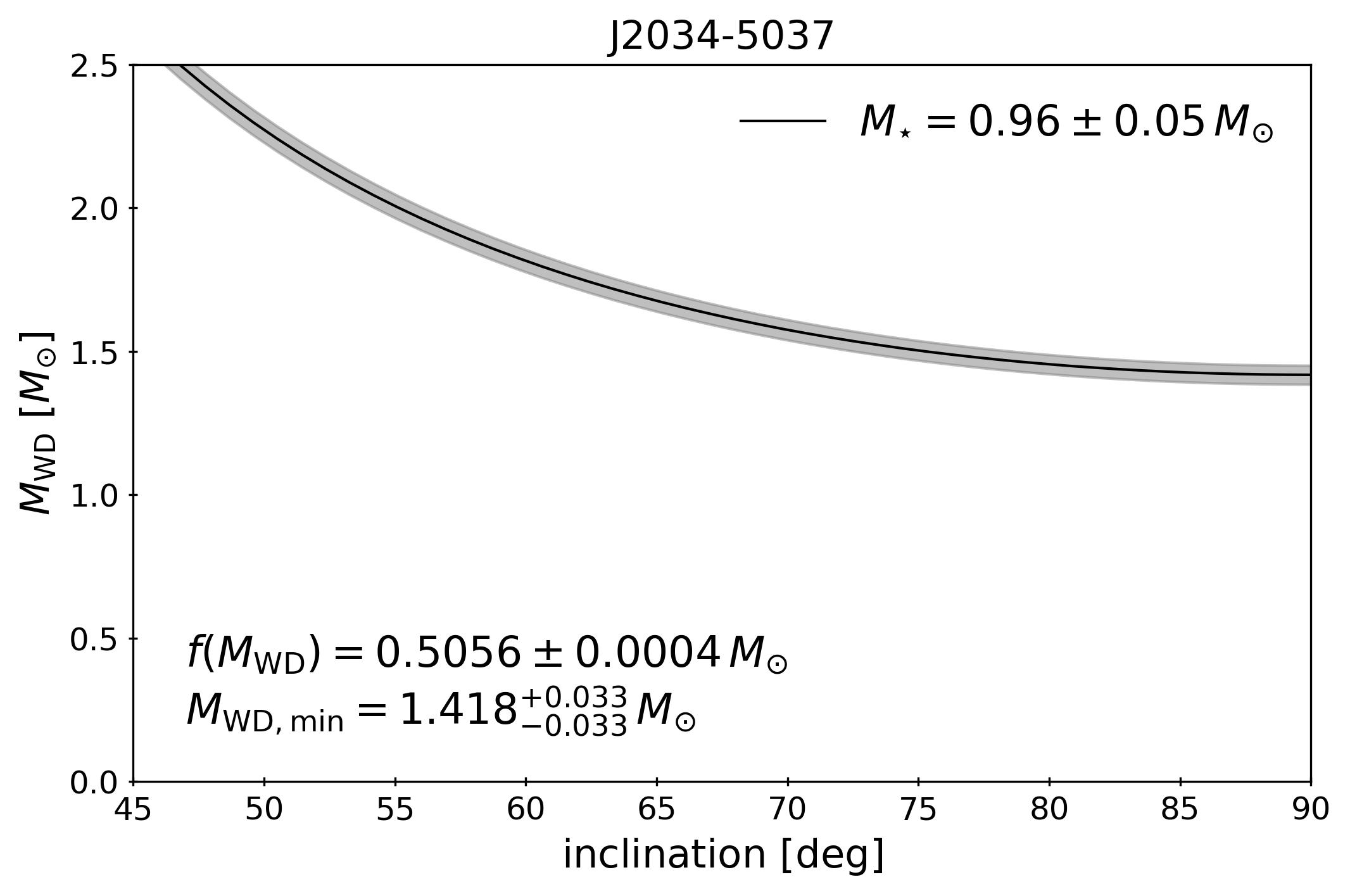}
     \includegraphics[width=0.33\textwidth]{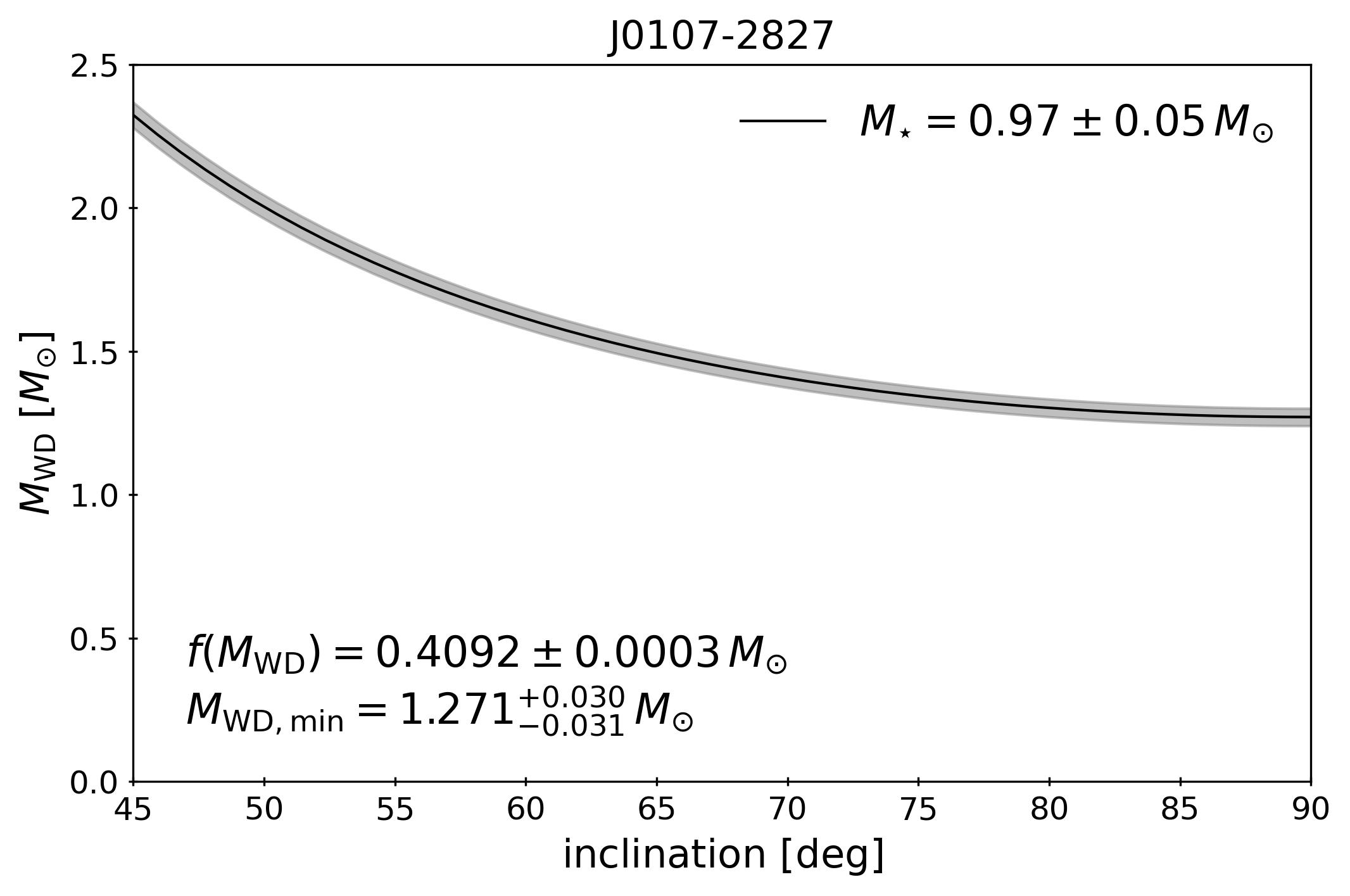}
     \caption{Plots of the implied WD mass as a function of inclination, as described in Section \ref{ssec:wd_masses}. For J1314+3818, we fit the RVs and {\it Gaia} astrometry simultaneously (Section \ref{ssec:astrometric}), with the red cross indicating the best-fit $i$ and the corresponding $M_{\rm WD} \sim 1.324\,M_{\odot}$.}
     \label{fig:WD_mass_func}
\end{figure*}

The implied $M_{\rm WD}$ as a function of the inclination is shown in Figure \ref{fig:WD_mass_func}. We shade the regions for $M_{\star}$ values $\pm 0.05\,M_{\odot}$ above and below the best-fit value from the SED fitting (Section \ref{sssec:parallax_errs}). The minimum masses $M_{\rm WD, min}$ range from $1.244\pm0.027$ to $1.418\pm0.033\,M_{\odot}$. Given the uncertainties, these values are all consistent with masses close to, but just below, the Chandrasekhar limit of $\sim 1.4\,M_{\odot}$. 

For J1314+3818, we obtain an inclination constraint from astrometry (Section~\ref{ssec:astrometric}) and thus a precise value for $M_{\rm WD}$ of $1.324 \pm 0.037\,M_{\odot}$, as opposed to just a lower limit. This point is shown as a red cross on the plot of the $M_{\rm WD}(i)$ for this object in Figure \ref{fig:RV_curves}. 

The inferred WD masses are summarized in Table \ref{tab:WD_masses}. At the time of writing, these WDs are among the most massive WDs known \citep[e.g.][]{Hermes2013ApJL, Curd2017MNRAS, Cognard2017ApJ, Hollands2020NatAs, Caiazzo2021Natur, Miller2022ApJL}, if they are indeed WDs (Section~\ref{ssec:alternatives}). 
We note that most other ultramassive WD candidates have mass estimates that depend on WD cooling models and mass-radius relations, which are uncertain for the most massive WDs \citep[e.g.][]{Camisassa2019A&A, Schwab2021ApJ}. Meanwhile, our measurements \citep[and similarly, those of][]{Cognard2017ApJ} provide fairly robust constraints on the mass with minimal assumptions about the WD itself (though there is still dependence on the stellar models used to infer the mass of the main-sequence companions). 

\begin{table}
    \centering
    \begin{tabular}{c|c}
        \hline
         name & M$_{\rm WD, min}$ [M$_{\odot}$] \\
         \hline
         J2117+0332 & 1.244$\pm$0.027 \\
         J1111+5515 & 1.367$\pm$0.029 \\
         J1314+3818 & 1.324$\pm$0.037 *  \\
         J2034-5037 & 1.418$\pm$0.033 \\
         J0107-2827 & 1.271$\pm$0.031 \\
         \hline
    \end{tabular}
    \caption{Minimum white dwarf masses using orbital parameters from the RV fitting and $M_{\star}$ from the SED fitting. J1314+3818 is marked in asterisk (*) to indicate that this value is not a lower bound, but the actual value given the inclination constraint from its astrometric solution (Table \ref{tab:rv_astro}).}
    \label{tab:WD_masses}
\end{table}

\subsection{Joint astrometric + RV fitting} \label{ssec:astrometric}

% 1522897482203494784 : RV + Astronometry, corner plot
One of our targets, J1314+3818, has a \textit{Gaia} AstroSpectroSB1 solution, meaning that the {\it Gaia} astrometry and RVs were fit with a combined orbital model. This model has 15 parameters: ra, dec, parallax, pmra, pmdec, a\_thiele\_innes, b\_thiele\_innes, f\_thiele\_innes, g\_thiele\_innes, c\_thiele\_innes, h\_thiele\_innes, center\_of\_mass\_velocity, eccentricity, period, t\_periastron \citep{Halbwachs2023A&A, GaiaCollaboration2023A&A}.  The Thiele-Innes elements A, B, F, and G describe the astrometric orbit of the photocenter and are transformations of the Campbell elements. The C and H elements are constrained from the {\it Gaia} RVs of the MS star. In the case of a dark companion, the photocenter simply traces the MS star. 

We fit our RVs and the {\it Gaia} constraints simultaneously, using the likelihood described by \citet{El-Badry2023MNRAS} which we briefly summarize here. \textit{Gaia} stores the correlation matrix of the parameters in a vector \texttt{corr\_vec}, from which, along with the errors on the parameters, we can construct a covariance matrix. We then construct a log-likelihood function that is a sum of two terms: one that compares the predicted astrometric parameters and all Thiele-Innes coefficients to the {\it Gaia} constraints, and one that compares the measured and predicted RVs (Equation~\ref{eqn:logL}).

The model we fit has 14 free parameters: ra, dec, parallax, pmra, pmdec, eccentricity $e$, inclination $i$, angle of the ascending node $\Omega$, argument of periastron $\omega$, periastron time $T_p$, center-of-mass RV $\gamma$, companion mass $M_{\rm WD}$, and luminous star mass $M_{\star}$. For $M_{\star}$, we set a Gaussian constraint based on its best fit value and error obtained from the SED fitting (Table \ref{tab:sed_params}). The resulting parameters can be found in Table \ref{tab:rv_astro} and the plots are shown in Figure \ref{fig:RV_curves}. We find $i \approx 99.9^{\circ}$ and $M_{\rm WD} = 1.324 \pm 0.037\,M_{\odot}$. 

\begin{table}
    \centering
    \begin{tabular}{c|cc}
         \hline
          & Astrometry Only & Astrometry + RV \\
         \hline
         RA [deg] & 198.517 $\pm$ 0.023 & 198.532 $\pm$ 0.010 \\
         Dec [deg] & 38.299 $\pm$ 0.013 & 38.307 $\pm$ 0.012 \\
         $\pi$ [mas] & 12.446 $\pm$ 0.015 & 12.447 $\pm$ 0.015 \\
         PMRA [mas/yr] & 129.523 $\pm$ 0.013 & 129.526 $\pm$ 0.011 \\
         PMDEC [mas/yr] & -224.216 $\pm$ 0.013 & -224.217 $\pm$ 0.013 \\
         $P_{\rm orb}$ [days] & 45.516 $\pm$ 0.005 & 45.519 $\pm$ 0.000 \\
         $e$ & 0.046 $\pm$ 0.006 & 0.0504 $\pm$ 0.0003 \\
         $i$ [deg] & 99.834 $\pm$ 0.370 & 99.945 $\pm$ 0.350 \\
         $\Omega$ [deg] & 86.017 $\pm$ 0.339 & 85.997 $\pm$ 0.344 \\
         $\omega$ [deg] & 99.219 $\pm$ 7.147 & 93.220 $\pm$ 0.307 \\
         $t_{\rm peri}$ [days] & -10.996 $\pm$ 0.916 & -11.763 $\pm$ 0.047 \\
         $\gamma$ [km/s] & 2.445 $\pm$ 0.209 & 2.811 $\pm$ 0.010 \\
         $M_{\rm WD}$ [M$_{\odot}$] & 1.325 $\pm$ 0.046 & 1.324 $\pm$ 0.037 \\
         $M_{\star}$ [M$_{\odot}$] & 0.713 $\pm$ 0.050 & 0.712 $\pm$ 0.049 \\
         \hline
    \end{tabular}
    \caption{Best-fit parameters of J1314+3818 from fitting just the astrometric solution, and when combined with the RVs.}
    \label{tab:rv_astro}
\end{table}

\subsection{Nature of the unseen companions} \label{ssec:alternatives}

Here we discuss whether the unseen companions could be objects other than WDs.

\subsubsection{MS binaries or triples}

A MS companion with a mass of $\sim 1.3\,M_{\odot}$ would dominate the SEDs of all the objects in our sample. In this case we would see two sets of lines in the spectra and changes in the composite line profiles with orbital phase. Since the spectra of our targets are all well-fit by single-star models, we can definitively rule out a single MS companion. 

A different possibility is that these systems are hierarchical triples consisting of a close inner binary of two $\sim 0.65\,M_{\odot}$ MS stars orbiting the primary \citep[e.g.][]{vandenHeuvel2020}. Together, the two would be dimmer than a single $1.3\,M_{\odot}$ MS star. We can estimate the contribution of such a binary to the overall SED in a similar way we do for the case of a WD in Appendix \ref{appendix:wd_sed}. We once again use \texttt{pytstelllibs} to generate an SED but for a $0.65\,M_{\odot}$ star on the MS. This mass roughly corresponds to a K7V star with a radius and temperature of $0.63 R_{\odot}$ and $4100$\,K respectively \citep[e.g.][]{Pecaut2013ApJS}. We can then calculate the ratio of the flux from the two stars to that of the single star which was fitted for in Section \ref{ssec:sed}. At 550 nm, the fractional flux contribution of such an inner triple would be  4.9, 2.6, 66.9, 13.24, and 5.2\%, respectively, for J2117+0332, J111+5515, J1314+3818, J2034-5037, and J0107-2827. In the infrared, at 3 $\mathrm{\mu m}$, the contribution would be larger, ranging from $\sim 13$ to $56 \%$ for four objects, with the exception of J1314+3818 where the inner binary would outshine the tertiary. Therefore, for J1314+3818, a hierarchical triple model is untenable. For the other objects, it is less obvious as there is a relatively small contribution at optical wavelengths meaning that any colour difference or spectral contribution is likely not enough to be distinguished from a single source. We also note that a WD+WD or WD+MS inner binary would similarly be dim in the optical and difficult to detect, but forming these in close orbits would be challenging from an evolutionary standpoint (given the size of the WD progenitor). The same challenges would apply for a massive WD formed from a WD+WD merger. Therefore, we do not consider these options further.

We also tested whether an inner binary's presence could be inferred from the SED fit. For each system, we constructed a ``mock triple'' SED by adding synthetic photometry for the inner binary to synthetic photometry for the best-fit single star model. We then fit this composite SED with a single-star model using MINEsweeper and check whether the residuals of the fit worsens significantly compared to those in Section \ref{ssec:sed}. We find that while the median residual does worsen slightly (at most by a factor of a few), the residuals of most photometric points still remain within $\lesssim 0.1$ mag. As expected, the exception is J1314+3818, where the residuals reach 0.2 mag. We conclude that the worst-case inner binaries could escape detection via a poor SED fit in all systems except J1314+3818. 

We next consider the possible periods of hypothetical inner binaries. There is a maximum period set by dynamical stability considerations: $P_{\rm out}/P_{\rm in} \gtrsim 5$ where $P_{\rm out}$ and $P_{\rm in}$ are the outer and inner orbital periods respectively \citep[we are also taking $e \sim 0$ and the ratio of the mass of the outer star to that of the inner binary $\sim 1$;][]{Mardling2001MNRAS, Tokovinin2014AJ}. Given that our objects have $P_{\rm orb} \sim 30$ days, this implies $P_{\rm in} \lesssim 6$ days. As for the minimum period, if the orbit of the inner binary is sufficiently tight, we may detect ellipsoidal variability due to tidal distortion of the inner components. We use the code PHysics Of Eclipsing BinariEs \citep[PHOEBE;][]{Prsa2005ApJ} to generate synthetic light curves of an inner binary of two K dwarfs for a range of periods from $\sim 0.25$ to 1 days. The amplitude of the ellipsoidal variability decreases with increasing period. We then add a fraction of the signal of these synthetic light curves to the observed light curves (described in Section \ref{ssec:lightcurves}). Given that an inner binary contributes $\lesssim 10$ \% of the total light (from above), we set this fraction to 0.1. We then generate periodograms of these  light curves \citep{AstropyCollaboration2022ApJ} and see whether or not we would be able to detect variability on half the inner binary's period. We find that with only $\sim 10$ \% of the light coming from the inner binary, ellipsoidal variability can only be distinguished from the noise for $P_{\rm inner} \lesssim 0.3$ days. Thus, the range of possible inner period is $\sim 0.3$ to 6 days. 

In summary, with the available observations, we cannot rule out a tight MS binary in four out of the five systems. However, we emphasize that there are very few hierarchical triple systems that have outer orbital periods below $\sim$ 1000 days \citep[this is not a selection effect, see][]{Tokovinin2014AJ}, while all our systems have $P_{\rm orb} < 50$ days. The few known compact hierarchical triples that have been found all have significantly more eccentric outer orbits than our objects, with values ranging from about 0.2 to 0.6. The only exceptions known are two triples in which the outer tertiary is a giant with a large radius, which likely circularized their orbits \citep{Rappaport2022MNRAS, Rappaport2023MNRAS}. All five of our systems have eccentricities close to zero and host MS stars in orbits that would not circularize in a Hubble time. This fact is easily understood if the companions are WDs -- the binaries would have been (partially) circularized when the WD progenitor was a red giant. If the companions were tight MS binaries, there would be no reason to expect circular outer orbits, and it would be very improbable for all 5 systems to have $e < 0.1$ by chance. 

\subsubsection{Neutron stars}

As we report minimum masses that are very close to the Chandrasekhar limit (Table \ref{tab:WD_masses}), we also consider the possibility that the companions are neutron stars (NSs). However, NSs are  expected to be born with natal kicks which drive their orbits to be eccentric \citep[e.g.][]{Hills1983ApJ, Colpi2002ApJ, Podsiadlowski2005ASPC}. Thus, we must consider formation mechanisms which can explain the near zero eccentricities of our objects (Table \ref{tab:rv_fitting}). 

In the case of no natal kick and spherically symmetric mass loss forming the NS \citep[e.g.][]{Blaauw1961}, the eccentricity acquired (taking an initial eccentricity of zero) is given by
\begin{equation}
    e = \frac{\Delta m}{m_c + m_2}
\end{equation}
where $\Delta m$ is the mass lost, $m_c$ is the remaining core/NS mass, and $m_2$ is the mass of the companion \citep[][]{Hills1983ApJ}. In the case where a $8\,M_{\odot}$ star explodes by a core-collapse supernova (SN) to form a $1.3\,M_{\odot}$ NS around a $1\,M_{\odot}$ MS companion, we see that $e > 1$ and the system will be unbound. In reality, the massive progenitor may lose a significant amount of mass through winds or binary interactins prior to the explosion in which case the eccentricity will be smaller and may allow the binary to survive, though likely in an eccentric orbit. Even in the case of an ultra-stripped SN explosion with $\sim 0.3\,M_{\odot}$ of ejecta \citep{De2018Sci, Yao2020ApJ}, we expect $e\sim 0.1$, significantly larger than the majority of our systems (though even lower ejecta masses are possible; \citealt{Tauris2013ApJL, Tauris2015MNRAS}). Moreover, if the SN is asymmetric (in its ejecta or neutrino emission), a strong kick can be imparted on the NS, which will likely result in large eccentricities, if it does not unbind the NS \citep[e.g.][]{Fryer1999ApJ, Tauris1999MNRAS}. Thus, a NS formed in this way is unlikely to be found in very circular orbits like our targets. 
Alternatively, a NS may form from a massive WD accreting up to the Chandrasekhar limit through accretion-induced collapse %(AIC) 
\citep[for a recent review on this topic, see ][]{Wang2020RAA}. Here, the ejecta masses are expected to be significantly smaller, though quite uncertain, with values ranging from $1\times 10^{-3}$ to $0.05\,M_{\odot}$ \citep{Darbha2010MNRAS, Fryer1999ApJ}. Such ejecta masses could correspond to $e \lesssim 0.02$ which are consistent with the eccentricities of some of our objects. However, it is difficult to explain how the progenitor WD would have accreted the necessary mass to begin with. Our systems all contain MS star companions which do not have strong winds, so there should be no significant wind accretion. %\citep{10.1088/978-0-7503-1278-3ch15}. 
Moreover, our objects are in orbits that are too wide for there to have been MT from the MS star through RLOF. Thus, while accretion-induced collapse could produce NSs in circular orbits, it struggles to do so in our systems where there are no obvious MT mechanisms. 

Therefore, we conclude that these alternative scenarios are improbable (though we emphasize that they are possible) and we proceed under the assumption that the unseen companions are WDs.

\section{Comparison to other binary populations} \label{sec:comparison}

Here  we compare the properties of our targets to other related classes of binaries, including other WD+MS PCEBs and WD + millisecond pulsar (MSP) binaries. 

\subsection{Literature PCEBs} \label{ssec:lit_pcebs}

%Plot of M2 vs Porb for these 5 objects compared to the SDSS PCEBs and IK Peg
% - include the prediction of Porb vs M2 from stable mass transfer to show these can’t have formed that way

The Sloan Digital Sky Survey \citep[SDSS;][]{Abazajian2009ApJS} detected large numbers of close WD+MS binaries. \citet{Rebassa-Mansergas2007MNRAS} identified 37 new PCEBs from the SDSS PCEB survey (described in Section \ref{ssec:other_surveys}), and combining this with 25 that were previously known, \citet{Zorotovic2010A&A} compiled a total of 62 PCEB systems. In addition, the ``white dwarf binary pathways survey'' also identified several PCEBs with AFGK companions via UV excess and RV variability detected from the RAVE and LAMOST surveys \citep[diamond markers;][]{Hernandez2021MNRAS, Hernandez2022MNRAS_vi, Hernandez2022MNRAS}. In Figure \ref{fig:avMwd}, we plot the minimum orbital separation $a_{\rm peri}$ against  $M_{\rm WD}$ for these literature PCEBs, the five objects from our sample, and several self-lensing WD+MS binaries discovered by the {\it Kepler} survey \citep[SLBs + KOI-3278;][]{Kawahara2018AJ, Kruse2014Sci}. With the exception of J1314+3818 where the precise value of $M_{\rm WD}$ was obtained using astrometry (Section \ref{ssec:astrometric}), we have plotted $M_{\rm WD, min}$ for our objects which is indicated with arrows. The colours of the points represent the mass of the luminous (MS) companion, $M_{\star}$. The gray dashed line comes from the $P_{\rm orb} - M_{\rm WD}$ relation derived in \citet{Rappaport1995MNRAS} for stable MT (with a spread in orbital period of a factor of $\sim 2.4$), where $P_{\rm orb}$ has been converted to separation assuming a $1\,M_{\odot}$ MS star and circular ($e = 0$) orbit (assuming instead a $0.1\,M_{\odot}$ M dwarf star only shifts this downwards by a small amount). The fact that all of these objects lie below this relation means that they are unlikely to have formed through stable RLOF, with the possible exception of the SLBs, which are not too far below the relation.

The blue line in Figure~\ref{fig:avMwd} shows the maximum radius of the WD progenitor, {\it if it evolved in isolation}. Binaries located below this line must have interacted at some point in their evolution. This approximate relation was obtained by first calculating the progenitor mass $M_{\rm init}$ using the WD Initial-Final Mass Relation (IFMR) derived in \citet{Williams2009ApJ}: $M_{\rm final} = 0.339 + 0.129 M_{\rm init}$, then generating MIST evolutionary tracks \citep{Dotter2016ApJS, Choi2016ApJ} to identify the maximum radius reached by a star with a given $M_{\rm init}$. Note that this is a conservative limit since the RLOF would begin before the giant touches the companion. 

\subsubsection{A population of PCEBs in wide orbits?}

IK\,Peg was previously isolated in its region of the $P_{\rm orb}-M_{\rm WD}$ parameter space: being in a wider orbit with a period of 22 days and hosting a more massive WD of $\sim 1.2\,M_{\odot}$ \citep{Wonnacott1993} than the vast majority of SDSS PCEBs. Our five targets fall in the same region as IK\,Peg. Their current orbits are far too tight for the binaries to have escaped interaction when the WD progenitors were red giants or AGB stars, strongly suggesting that these objects are indeed PCEBs. As we show in Section~\ref{sec:mesa}, the current orbits can only be understood as an outcome of CEE if additional energy sources (besides liberated orbital energy) helped unbind the common envelope. 

\begin{figure*}
    \centering
    \includegraphics[width=\textwidth]{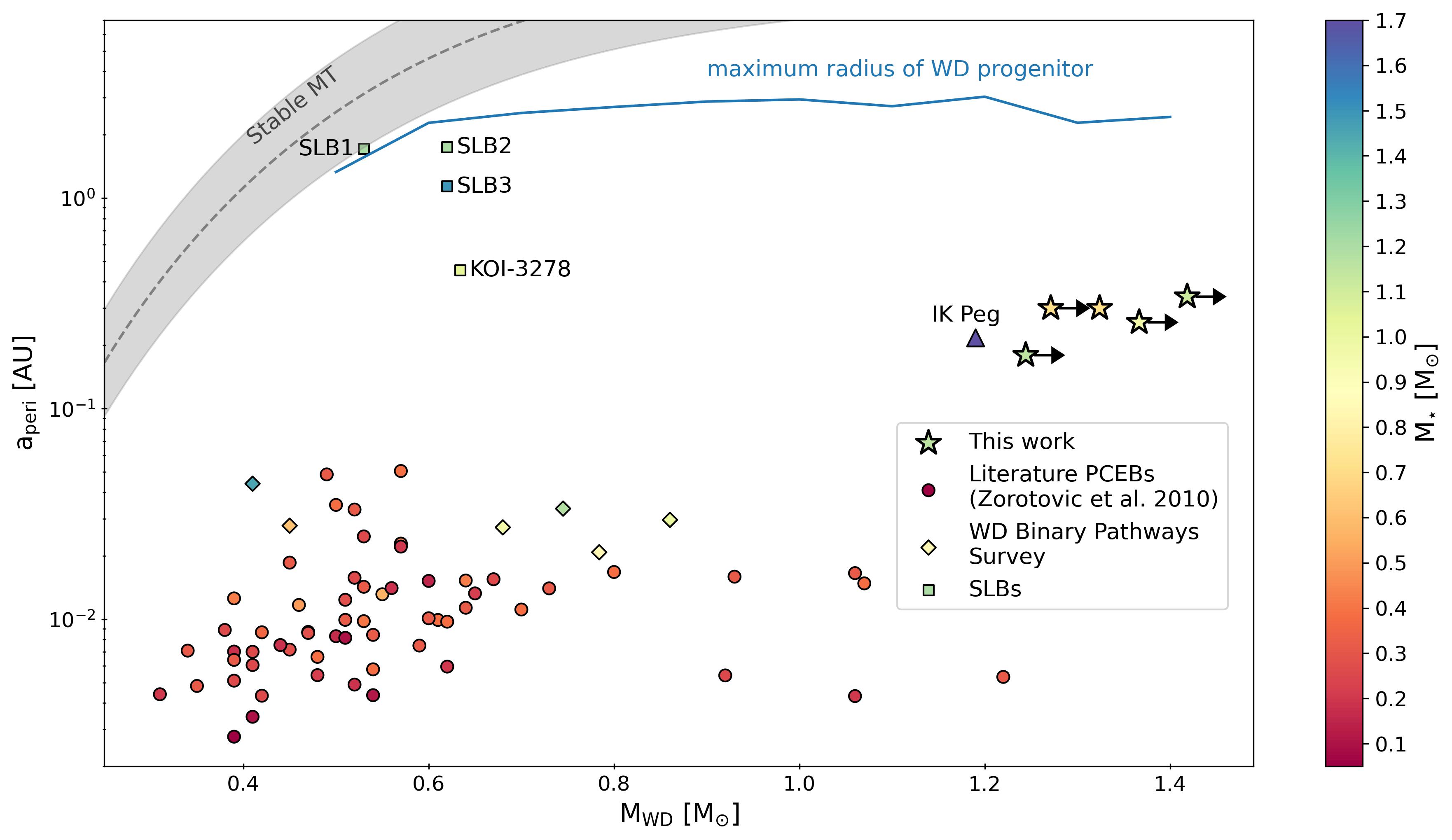}
    \caption{Periastron separation, $a_{\rm peri}$, vs. WD mass, $M_{\rm WD}$, for a sample of literature PCEBs (circle markers; compiled by \citealt{Zorotovic2010A&A}) and the five objects from this work (star markers; the arrows indicate lower limits). IK\,Peg is distinguished from the other known PCEBs with a triangle marker as it lies very close to our objects. We also plot the PCEBs from the ``white dwarf binary pathways survey" \citep[diamond markers;][]{Hernandez2021MNRAS, Hernandez2022MNRAS_vi, Hernandez2022MNRAS}. Finally, we plot self-lensing binaries (SLBs) discovered by \citet{Kawahara2018AJ} as well as  KOI-3278 \citep[][also a SLB]{Kruse2014Sci}, which were all detected by {\it Kepler} (square markers). The colours of the points represents the masses of the luminous MS companions. The dashed gray shows the prediction for stable MT from \citet{Rappaport1995MNRAS} and the blue line indicates the maximum radius reached by the WD progenitor for a range of WD masses. }
    \label{fig:avMwd}
\end{figure*}

The SLBs occupy a different isolated region in this space, with normal WD masses but at separations even wider than our systems and IK\,Peg. Like our targets, they contain solar-type MS stars, which are more massive than the M dwarfs in the SDSS PCEB sample. \citet{Kruse2014Sci} initially interpreted KOI-3278 as a ``normal'' PCEB, but \citet{Zorotovic2014A&A} subsequently showed that the system's wide orbit requires an extra source of energy, beyond orbital energy, to contribute to the CE ejection. The three SLBs identified by \citet{Kawahara2018AJ} have even wider orbits than KOI-3278. Those authors interpreted the systems as having formed through stable MT, but Figure~\ref{fig:avMwd} shows that SLB 2 and 3 fall well below the \citet{Rappaport1995MNRAS} prediction. Formation through stable MT thus seems tenable only if these WDs have significantly overestimated masses.

We also distinguish the objects discovered by \citet{Hernandez2021MNRAS, Hernandez2022MNRAS_vi, Hernandez2022MNRAS} (plotted with diamond markers) from the SDSS PCEBs as they host higher-mass ($\sim 1\,M_{\odot}$) MS stars. Compared to our objects, these have shorter orbital periods ($P_{\rm orb} \sim 1-2$ days) and can therefore be explained with just the liberated orbital energy, without needing to invoke additional sources. This tells us that having an intermediate-mass MS star as a companion does not necessarily lead to a wide post-CE orbit.  

We note that while Figure \ref{fig:avMwd} may suggest that there are two distinct groups of PCEBs in wide orbits with different WD masses (self-lensing binaries vs. IK-Peg analogs), this is not necessarily the case. We remind the reader that here we specifically targeted very massive companions which would at least partly explain why we did not find any that are less massive. A search for more objects located in currently sparse regions on the plot would be useful.

% Plot eccentricity vs Porb and compare to the pulsar + white dwarf binaries

\subsection{Eccentricities and comparison to MSP binaries} \label{ssec:MSPs}
\begin{figure}
    \centering
    \includegraphics[width=\columnwidth]{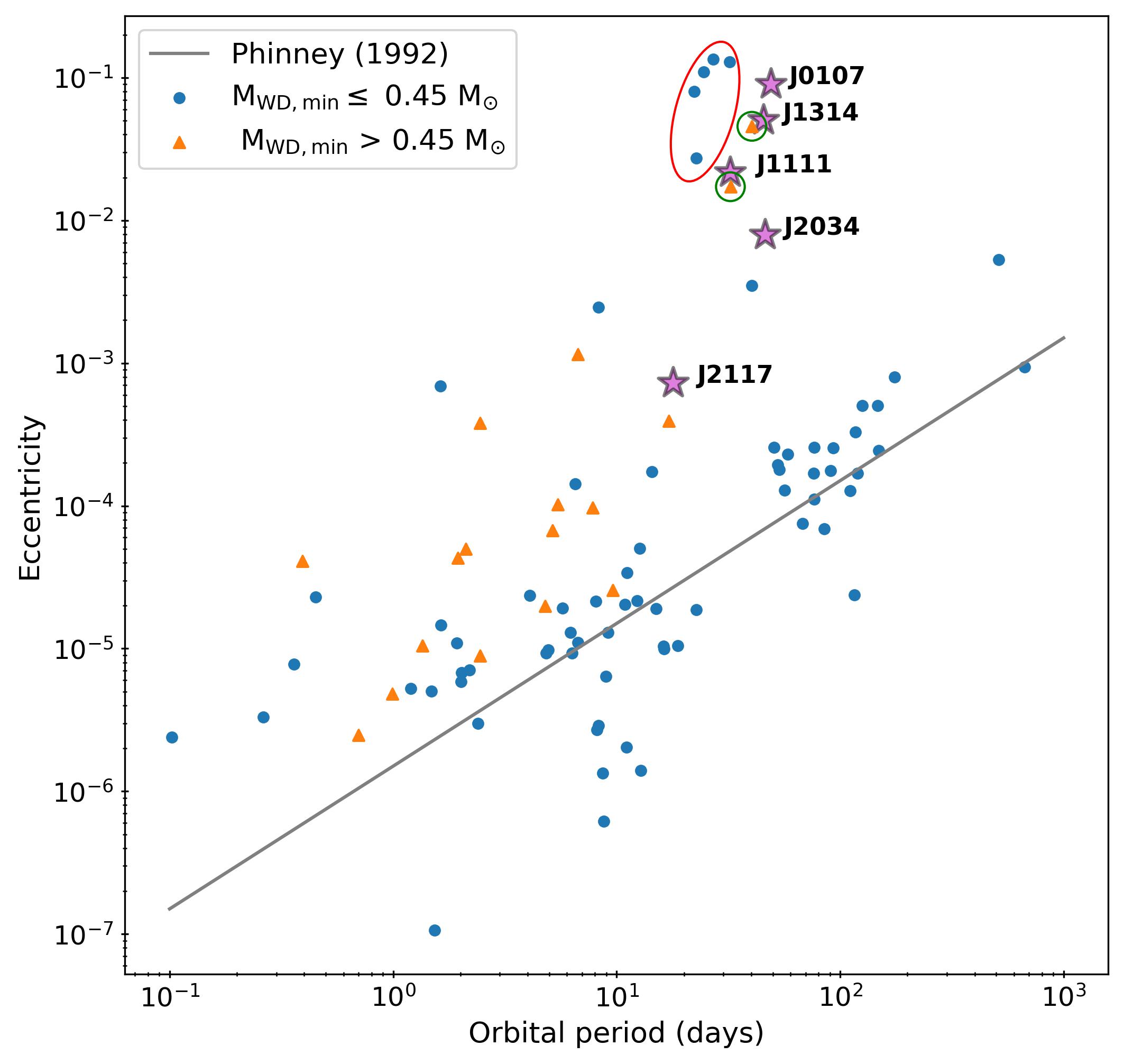}
    \caption{Eccentricity vs. orbital period. We plot the sample of MSP + WD binaries mainly from the ATNF catalogue (\citealt{Manchester2005AJ}, with a few others, compiled by \citealt{Hui2018ApJ}), differentiating between those with minimum WD mass above and below $0.45\,M_{\odot}$ (orange triangles and blue circles, respectively). We show our objects with magenta stars. The gray line is the theoretical relation for MSP + He WD binaries formed through stable MT derived by \citet{Phinney1992RSPTA}. The points circled in red are eMSPs \citep[e.g., see ][for a description of the five plotted here]{Stovall2019ApJ}. We also circle in green two binaries with massive CO WDs with large eccentricities \citep{Lorimer2015MNRAS, Berezina2017MNRAS}.}
    \label{fig:PvEcc}
\end{figure}

On Figure \ref{fig:PvEcc}, we plot the eccentricity $e$ against $P_{\rm orb}$ and compare our objects to MSP + WD binaries. We plot objects primarily from the Australia Telescope National Facility (ATNF) catalogue \citep{Manchester2005AJ} (taking the version analyzed by \citealt{Hui2018ApJ}) and distinguish those with minimum WD companion masses above and below $0.45\,M_{\odot}$ (this is the approximate upper limit to the mass of a He WD). We also plot the theoretical relation derived by \citet{Phinney1992RSPTA} for MSP pulsars with He WDs formed through stable MT.

MSPs form through ``recycling'', where an old NS is spun up to short periods by the transfer of mass and angular momentum from a companion \citep[e.g.][]{Alpar1982Natur, Radhakrishnan1982CSci, Bhattacharya1991}. Tides are almost expected to circularize MSP + WD binaries, but a very low orbital eccentricity remains %small amount of orbital eccentricity survives 
because convection in the WD progenitor produces a time-varying quadrupole moment, leading to perturbations and eccentricity excursions that are larger in longer-period systems, which hosted larger giants \citep{Phinney1992RSPTA}. To date, the period--eccentricity relation has mainly been tested with MSP+WD binaries, because their eccentricities can be easily measured with high precision. However, a similar process should operate in MS+WD binaries, if MT occurs over a long enough period for tidal circularization to occur. 

Figure~\ref{fig:PvEcc} shows that in general, the MSP binaries with more massive (CO/ONeMg) WDs tend to have higher eccentricities at fixed period than those with low mass (He) WDs. The standard interpretation  is that the systems with He WDs formed via stable MT, while those with more massive WDs formed through CEE \citep[e.g.][]{vandenHeuvel1994, Tauris2012}. Although NS + red giant orbits are expected to be circularized prior to the onset of MT, eccentricity is produced during the dynamical plunge-in phase of CEE \citep[e.g.][]{Ivanova2013A&ARv}, and there is likely insufficient time for this eccentricity to be fully damped between the end of CEE and the formation of the WD \citep[e.g.][]{Glanz2021MNRAS}. 

The objects in our sample have periods and eccentricities similar to the longest-period MSP + CO/ONeMg binaries, perhaps pointing to a similar formation history. In particular, the MSP binaries J1727-2946 \citep{Lorimer2015MNRAS} and J2045+3633 \citep{Berezina2017MNRAS} (circled in green) are located very close to our objects in Figure \ref{fig:PvEcc}. Both of these systems contain mildly recycled pulsars (with spin periods of 27 and 32 ms) and massive WDs ($M_{\rm WD} > 0.8\,M_{\odot}$), and they have eccentricities of 0.045 and 0.017 at orbit periods of 40 and 32 days. A common envelope origin has also been proposed for J2045+3633 \citep{McKee2020MNRAS}. % Meanwhile, the origin of their large eccentricities remain unclear (discussed below). 
There are also stable MT scenarios that have been proposed to explain the formation of MSP + CO/ONeMg WD binaries \citep[e.g.][]{Tauris2000}, which would likely predict lower eccentricities. We refer readers to \citet{Tauris2011ASPC} for a concise overview of this topic. 

It is worth mentioning that several systems with low-mass WDs in anomalously eccentric orbits have been discovered, called eccentric MSPs \citep[eMSPs; e.g.][]{Bailes2010}. These points are circled in red in Figure \ref{fig:PvEcc}. Interestingly, these eMSPs occupy a narrow range in orbital periods with eccentricities that are comparable to the two eccentric MSP + CO WD binaries \citep[as pointed out by][]{Berezina2017MNRAS}. This is unexpected, as binaries with He WDs are thought to have formed through distinct evolutionary paths, involving long periods of stable MT in low mass X-ray binaries \citep{Tauris2011ASPC}. There have been multiple proposed mechanisms to explain the large eccentricities, some of which may be applicable to one or more of the systems just discussed (MSP + He WD, MSP + CO WD, MS + WD PCEB). These include MSPs being directly formed from the accretion-induced collapse of a super-Chandrasekhar mass ONeMg WD \citep{Freire2014MNRAS}, interaction with a circumbinary %(CB) 
disk formed from material lost from the WD during H shell flashes \citep{Antoniadis2014ApJL}, or a circumbinary disk formed from the ejected envelope after a CE phase \citep{Dermine2013A&A}. %Using smoothed particle hydrodynamical simulations, \citet{Glanz2021MNRAS} also found that if the initial binary is highly eccentric, the CE phase can produce PCEBs with final eccentricities up to 0.18. 

\subsection{Other related populations} \label{ssec:related_pops}

Several other binary populations have been identified as likely products of CEE but have been largely neglected by works inferring CEE parameters from observed PCEB populations. We discuss them and their possible relation to the objects in our sample below.
 
First, the ``post-AGB'' binaries \citep{2003ARA&A..41..391V, Oomen2018A&A}. Post-AGB stars are short-lived transitional objects that have lost most all of their envelopes during the AGB phase and are transitioning toward hotter temperatures at constant luminosity on their way to the WD cooling track. Most of the known post-AGB stars have temperatures of $T_{\rm eff}=5000-8000$\,K and radii of $20-50\,R_{\odot}$. About 50 post-AGB stars are known to be in binaries, with periods ranging from $\sim$100 to a few thousand days. These periods are too short for the binaries to have avoided interaction during the AGB phase, and thus post-AGB binaries are generally modeled as having formed through CEE \citep[e.g.][]{Izzard2018}. Like the objects in our sample, their wide orbits would require very efficient envelope ejection to explain.

It has also been established that a significant fraction \citep[a lower limit of $10 - 20\%$;][]{Bond2000ASPC, Miszalski2009A&A} of the central stars of planetary nebulae (CSPNe) are in  binaries \citep[for a recent review, see][]{Jones2017NatAs}. The majority of the known systems were discovered by their photometric variability and have orbital periods $\lesssim 1$ d \citep[e.g.][]{Miszalski2009A&A, Jacoby2021MNRAS}. Given the short lifetimes of planetary nebulae, these should have undergone the CEE very recently with little subsequent evolution. Post-AGB stars are immediate progenitors of planetary nebulae, so the striking mismatch between the long periods of post-AGB binaries and short periods of binary CSPNe is puzzling. We suspect that it owes mostly to observational selection effects: post-AGB stars are too large and puffy to fit inside short-period orbits, while longer-period binary CSPNe are difficult to detect. If this explanation holds, a significant population of intermediate- and long-period binary CSPNe should still await detection. Indeed, a few binary CSPNe have been found with longer orbital periods of $\sim 10 - 30$ d \citep[e.g.][]{Brown2019MNRAS, Jacoby2021MNRAS}, comparable to those of IK Peg and our five systems, and with long-term RV monitoring, a few with orbital periods of thousands of days have also been detected \citealt[][]{VanWinckel2014A&A, Jones2017A&A}.

Another related class of binaries is the barium stars, which are red giants found in binaries that show an enhancement in s-process elements, such as barium. Their peculiar abundances are thought to be the result of the star accreting material from an AGB  star companion though winds \citep{Boffin1988A&A} or RLOF \citep{McClure1983ApJ}. The AGB star then evolves into a WD, forming a PN in the process and leaving behind the barium stars in relatively wide orbits at the centers of PNe. A few such systems have indeed been found \citep[e.g.][]{Miszalski2012MNRAS, Miszalski2013MNRAS}, while many more have been found with similar orbital periods that are not in PNe \citep[e.g.][]{Jorissen2019A&A}. Most known barium stars have periods of 200-5000\,d: comparable to the post-AGB binaries, but wider than the five objects in our sample. Here too, the fact that the polluted stars are giants likely introduces a significant selection bias in favor of long-period systems, and short-period analogs hosting MS stars  have recently been identified \citep{Roulston2021}.

Finally, a subset of the low-eccentricity binaries with periods of $\sim 200 - 1500$ d discovered via phase modulation of $\delta$ scuti pulsations \citep{Murphy2014MNRAS, Murphy2018} are likely to contain WD companions, many of which likely formed through a similar channel to the barium and post-AGB stars. Since the amplitude of phase modulations depends on the physical size of the orbit, selection effects in this sample likely also favor long periods. 

To summarize, the orbital periods of the binaries in our sample fall between the tight binaries that have previously been used to constrain CEE models (WD+MS PCEBs and binary CSPNe), and the long-period post-AGB and barium star binaries, which may have formed through either CEE or wind accretion. Given the complex selection effects affecting all the observed post-interaction binaries, more work is required to understand how these populations are related and to infer the intrinsic post-CEE period distribution.

%The properties of these stars provide constraints on the mass transfer histories of MS+WD binaries.

\section{Feasibility of formation through common envelope evolution} \label{sec:mesa}

%some plots characterizing how much energy it would take to unbind envelope (details TBD)
%can we infer an effective “alpha” for these systems?

To test whether our targets could have formed via CEE, we ran Modules for Experiments in Stellar Astrophysics (MESA) models \citep{Paxton2011ApJS, Paxton2013ApJS, Paxton2015ApJS, Paxton2018ApJS} of the progenitor star to the WD in our systems, evolving an intermediate-mass star up to the asymptotic giant branch (AGB). We emphasize that we are not using binary models to trace the evolution during the CE phase, which is beyond the capabilities of MESA. We are simply constructing a realistic model of the giant at the onset of mass transfer to calculate the energy budget as described below.

In the $\alpha$-formalism, the CEE ends when the loss of orbital energy from the spiral-in exceeds the binding energy of the envelope $E_{\rm bind}$, resulting in the envelope being ejected:
\begin{equation} \label{eqn:alpha_formalism}
    E_{\rm bind} = \alpha_{\rm CE} \left(-\frac{GM_{\rm WD} M_{\star}}{2a_f} + \frac{GM_{i} M_{\star}}{2a_i}\right)
\end{equation} 
where $M_{i}$ is the mass of the WD progenitor, $a_i$ is the initial separation at the onset of CE, $a_f$ is the separation at the end of the CE phase, and $\alpha_{\rm CE}$ is the fraction of the liberated orbital energy that goes into unbinding the envelope. Thus, $E_{\rm bind}$ determines the final separation of the system for a given initial separation. 

In the simplest case, $E_{\rm bind}$ is just the gravitational binding energy of the envelope. However, previous works using BPS have found that in this case, no values of $\alpha_{\rm CE} < 1$ can reproduce the relatively wide orbits of IK\,Peg and KOI-3278 \citep{Davis2010MNRAS, Zorotovic2010A&A, Zorotovic2014A&A, Parsons2023MNRAS}, which have comparable separations to our systems. Moreover, it is clear that additional energy exists within the stellar envelope, which can potentially help to unbind it. %This means that we must consider additional sources of energy that make the envelope easier to remove. 
Here, we consider the inclusion of internal energy which is defined in MESA as the sum of thermal and recombination energy \citep{Paxton2018ApJS}. This makes the binding energy less negative (possibly even positive), corresponding to an envelope that is less bound. Recombination of H and He can occur if there is some process (in this case, the binary interaction through the CE phase) that causes the envelope to expand and cool down, which will release energy \citep[e.g.][]{1968IAUS...34..396P, Ivanova2013A&ARv}. 

From initial-final mass relations of WDs \citep[e.g.][]{Williams2009ApJ, Cummings2018ApJ, El-Badry2018ApJL}, we expect the MS progenitor masses of our ultra-massive WDs to be in the range of $6 - 9\,M_{\odot}$. Thus, we run MESA models of a $7\,M_{\odot}$ star, following its evolution up to the AGB. Our inlists are based on those of \citet{Farmer2015ApJ} \citep[same wind and mixing prescriptions; For more details, we refer readers to Section 2 of ][]{Farmer2015ApJ}, although we have updated them for the more recent MESA version r22.05.1. We only run non-rotating models. 

\begin{figure*}
    \includegraphics[width=0.25\textwidth]{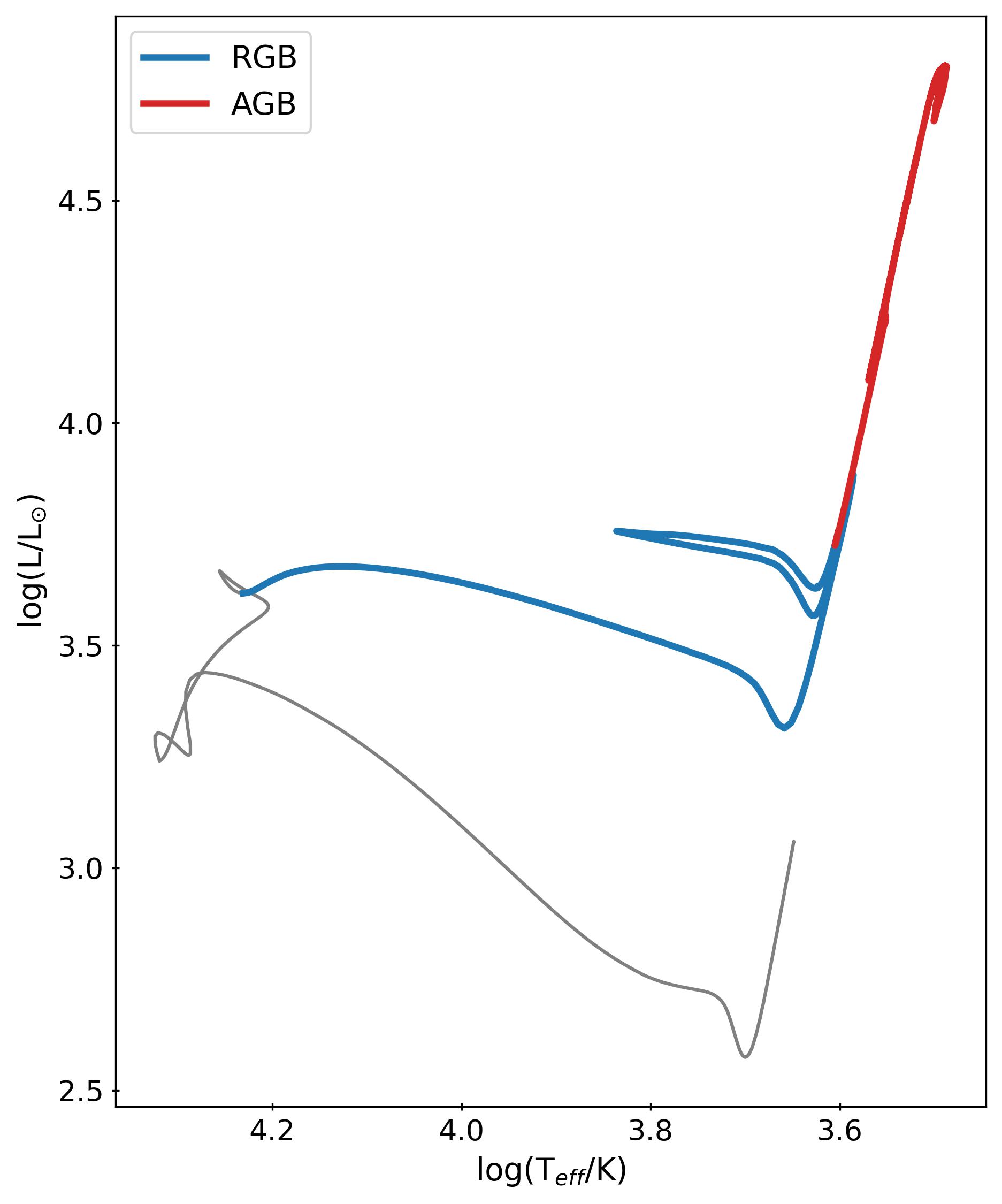}
    \includegraphics[width=0.72\textwidth]{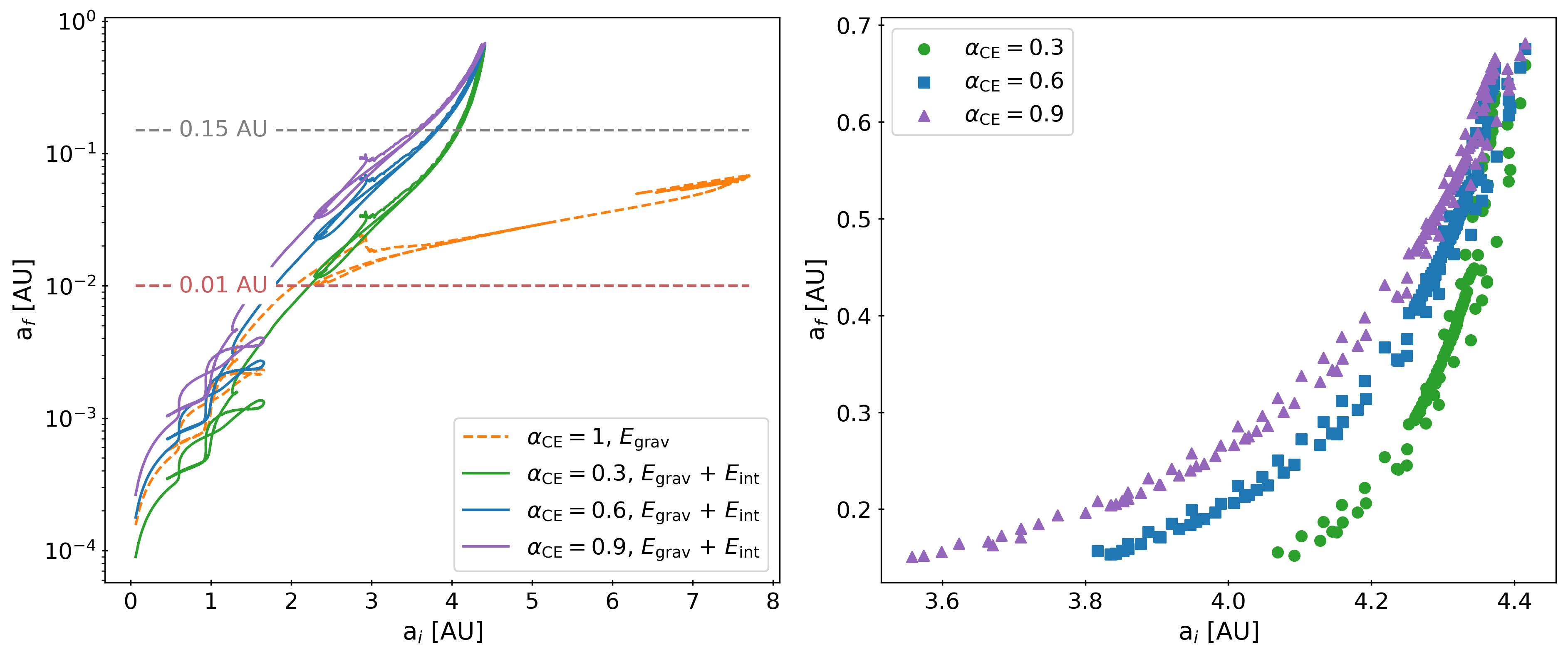}
    \caption{\textit{Left}: HR diagram showing the evolution of a $7\,M_{\odot}$ star starting from pre-MS to the AGB. The blue sections indicate what we refer to as the RGB (but which also includes the SGB) and the red sections represents the AGB. \textit{Center}: Plots of the final separation $a_f$ (i.e. birth period at the end of CEE) over a range of initial separations $a_i$. $a_i$ is taken to be the orbital semi-major axis when the giant (WD progenitor) fills its Roche lobe. We mark $a_f =$ 0.01 AU $\sim 2 R_{\odot}$ (red dashed line) below which the MS star would not fit in the orbit and a PCEB cannot form. The orange dashed line is the case where only the gravitational binding energy is considered and $\alpha_{\rm CE} = 1$. We see that in this case, no values of $a_i$ result in $a_f > 0.15$ AU (gray dashed line) which is approximately the minimum separation of our objects. The other three lines are the cases where all of the internal energy is added to the binding energy for $\alpha_{\rm CE} = 0.3$, 0.6, and 0.9. We see that these lines lie above the dashed line for some range of $a_i$. \textit{Right}: Zoom in on the region where $a_f > 0.15$ AU. We see that overall, $a_i \sim 3.5 - 4.4$ AU can result in the wide orbital separations of our systems.}
    \label{fig:mesa_af}
\end{figure*}

The evolution of the $7\,M_{\odot}$ model, from pre-main sequence to termination at the tip of the AGB, is shown on the HR diagram in the leftmost panel of Figure \ref{fig:mesa_af}. In the following, we consider progenitors on the red giant branch (RGB), including the sub-giant branch (SGB), and the AGB which are highlighted in blue and red on the diagram respectively. We define these phases following the convention used in the MIST project \citep[See Section 2.1 of ][]{Dotter2016ApJS}. We note that our model terminates before core carbon burning, but as the envelope binding energy becomes positive before this point (see below), our main conclusions should not be largely affected by this.

At each timestep, we calculate the binding energy of the envelope as a sum of the gravitational ($E_{\rm grav}$) and internal ($E_{\rm int}$) components:
\begin{align} \label{eqn:binding_energy}
    E_{\rm bind} &= E_{\rm grav} + E_{\rm int} \\
    &= \int_{M_{\rm core}}^{M_{\rm tot}} -\frac{G m}{r(m)} + U(m) dm
\end{align}
where $E_{\rm grav}$ and $E_{\rm int}$ correspond to the first and second parts of the integrand respectively, $m$ is the mass enclosed within a radius $r$, and $U(m)$ is the internal energy per unit mass. The integral is taken from the He core boundary to the surface of the star. We use the default definition of the He core in MESA which is where the hydrogen mass fraction $X_H < 0.1$ and helium mass fraction $X_{\rm He} > 0.1$. We have tested changing these boundaries to 0.01 and find no significant change. Here, we assume that all of the internal energy, both the thermal and recombination components, contributes to the envelope binding energy. We discuss the consequences of relaxing this assumption in the following section.%Appendix \ref{appendix:alpha_rec}. 

Using equation \ref{eqn:alpha_formalism}, we can solve for $a_f$ for Roche lobe overflow (RLOF) occurring at different points on the RGB and AGB. We calculate the initial separation $a_i$ using the Eggleton formula \citep{Eggleton1983ApJ}, assuming the giant fills its Roche lobe at the onset of the CEE. The mass of the WD remaining after the envelope ejection, $M_{\rm WD}$, is taken to be equal to the helium core mass of the giant, $M_{\rm core}$. This quantity grows from $\sim 1.2$ to $1.7\,M_{\odot}$ on the RGB and falls back down to $\sim 1.1\,M_{\odot}$ on the AGB, when some of the helium is mixed back into the envelope during second dredge-up, \citep[e.g.][]{Busso1999ARA&A}. We assume $M_{\star} = 1\,M_{\odot}$, which is close to the median value for our objects (Table \ref{tab:sed_params}). 

The predicted $a_f$ is shown for a range of $a_i$ in Figure \ref{fig:mesa_af}. In the central panel, the gray dashed line marks $a_f = 0.15$ AU, which is slightly smaller than the minimum $a_{\rm peri}$ of our objects at $\sim 0.18$ AU (Figure \ref{fig:avMwd}). The red dashed line marks $a_f = 0.01$ AU ($\sim 2\,R_{\odot}$), below which a $\sim 1\,M_{\odot}$ MS star cannot fit inside the orbit and thus a PCEB would not form (a merger, or perhaps stable MT of the MS onto the WD, may occur instead). 
%Firstly, we notice that the mass of the core does not grow monotonically but decreases while on the AGB. Specifically, there is a fall in the mass of the He layer of the core. This is likely the result of the second dredge-up where convection transports the central He to the surface \citep[e.g.][]{Busso1999ARA&A}. 
The orange line shows the case where we only consider the gravitational binding energy ($E_{\rm bind} = E_{\rm grav}$) and set $\alpha_{\rm CE} = 1$ (i.e. 100\% of the orbital energy loss goes into envelope ejection). We see that it never crosses the $a_f = 0.15$ AU mark, meaning that even in this optimistic case, there is not enough energy to unbind the envelope and produce orbits as wide as our observed systems. In the remaining three cases, we include the internal energy ($E_{\rm bind} = E_{\rm grav} + E_{\rm int}$, Equation \ref{eqn:binding_energy}) and let $\alpha_{\rm CE} = 0.3$ (the ``standard" value), 0.6, and 0.9. We see that for each case, there is a region where $a_f$ exceeds 0.15 AU. On the right panel, we zoom into this region and find $a_i$ ranges from $\sim$ 3.5 - 4.4 AU across all values of $\alpha_{\rm CE}$, with a narrower range for lower values of $\alpha_{\rm CE}$. 

We also ran a $6 M_{\odot}$ WD progenitor model and found qualitatively similar results but with the $a_i$ range for which $a_f$ > 0.15 AU shifted to $\sim 3.2 - 4.1$ AU. Given the simplified treatment here, these small quantitative differences should not be over-interpreted (and hence we do not explore them further) but it does tell us that there is likely a broader range of initial separations for which wide PCEBs can be formed than that implied by a single model.

In Figure~\ref{fig:mesa_af}, we exclude models in regions of parameter space where the $\alpha$-formalism does not make clear predictions, namely, those in which $E_{\rm bind} > 0$ (the envelope is unbound when recombination energy is included).  % and models in which $E_{\rm bind} < 0$ but $a_f > a_i$ (the orbit expands). 
This case is likely still relevant for producing wide PCEBs, but it is unclear what the final separation should be: if the MS star does not penetrate deep into the giant's envelope, it is unlikely to trigger the release of much recombination energy. For the model shown in Figure~\ref{fig:mesa_af}, these conditions are realized late in the AGB evolution, corresponding to radii of $\sim 520 - 888 R_{\odot}$ and initial separations of $\sim 4.4 - 7.7$AU.

\subsection{Efficiency of recombination energy} \label{ssec:alpha_rec}

The calculation above consider two kinds of internal energy: thermal energy and recombination energy. While thermal energy is commonly considered as an ``extra'' energy source, it is closely related to the gravitational potential energy through the virial theorem, and all the thermal energy should be included in the binding energy by default \citep[e.g.][]{Ivanova2013A&ARv}. The inclusion of recombination energy is more uncertain.

Some works have argued that most of the energy released by recombination will quickly be transported to the photosphere through radiation and/or convection and then radiated away \citep{Sabach2017MNRAS, Grichener2018MNRAS}. Meanwhile, \citet{Ivanova2018ApJL} found that such energy transport is inefficient in typical AGB stars and that recombination is in fact a significant source of additional energy. The effectiveness of recombination energy in widening the final orbit has also been explored in hydrodynamic simulations, with a range of results \citep[e.g.][]{Reichardt2020MNRAS, Gonzalez-Bolivar2022MNRAS}. Given this uncertainty, our previous assumption that all of the internal energy contributes to unbinding the envelope may be too optimistic.  

We can assess the sensitivity of our results to this assumption by splitting the internal energy into two components:
\begin{equation}
    E_{\rm bind} = E_{\rm grav} + \alpha_{\rm th} E_{\rm th} + \alpha_{\rm rec} E_{\rm rec}
\end{equation}
where $E_{\rm th}$ and $E_{\rm rec}$ are the thermal and recombination energies, and $\alpha_{\rm th}$ and $\alpha_{\rm rec}$ are the respective efficiencies. As described above, we set $\alpha_{\rm th} = 1$. 

MESA does not provide a simple way to individually track the thermal and recombination energies. We can, however, approximate the thermal energy using the ideal gas law, which is a reasonable approximation in the envelopes of AGB and RGB stars. In this case, the energy density is $(3/2)P/\rho$, where $P$ is the pressure and $\rho$ is the mass density. We subtract this from the total internal energy output by MESA to get an estimate of the recombination energy and see what value of $\alpha_{\rm rec}$ would be required to produce wide orbits. 

\begin{figure}
    \centering
    \includegraphics[width=\columnwidth]{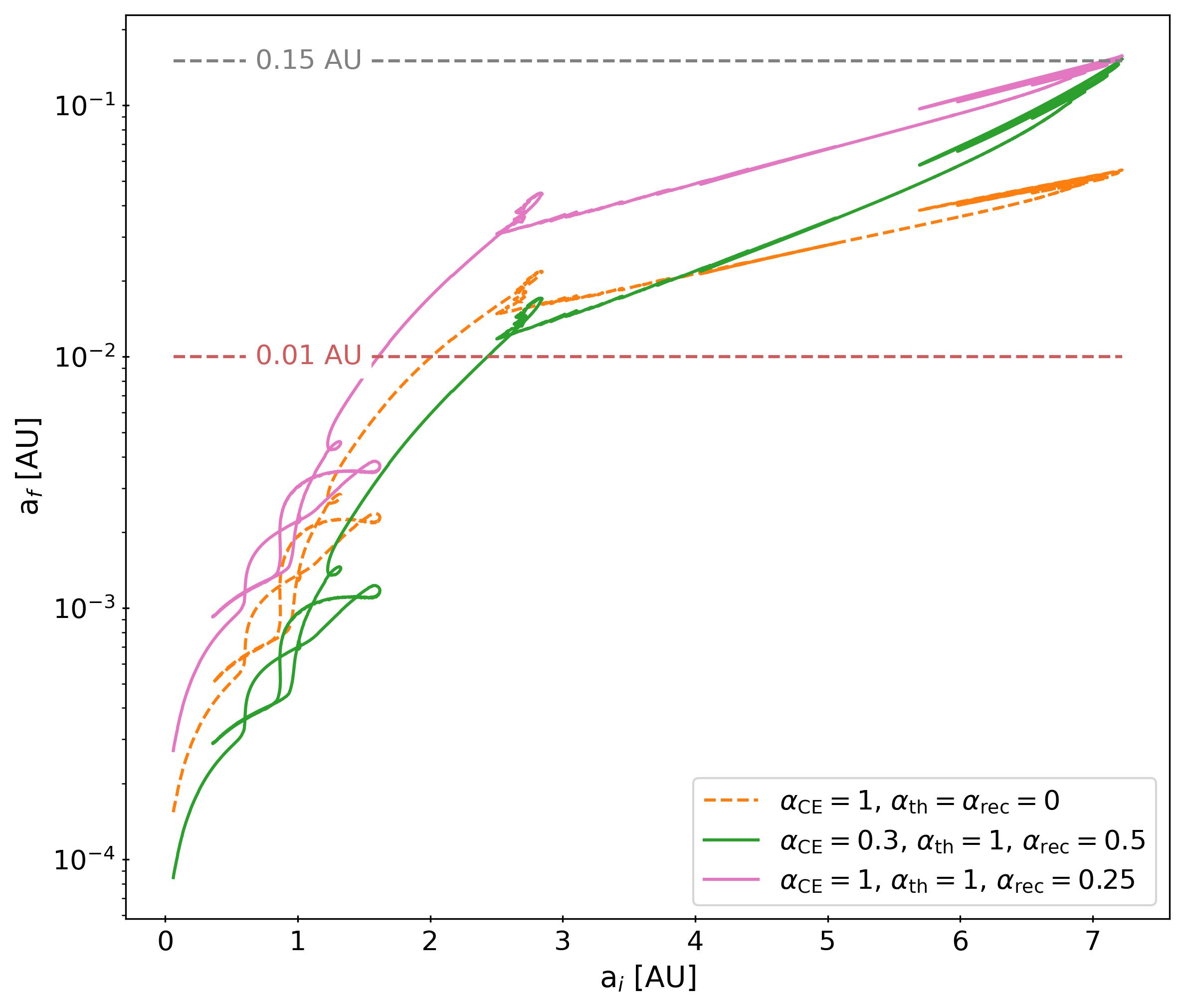}
    \caption{The same plot as in the central panel of Figure \ref{fig:mesa_af} but separating out the thermal and recombination components of the internal energy $E_{\rm int}$. As in Figure \ref{fig:mesa_af}, we plot the case where only $E_{\rm grav}$ is considered with $\alpha_{\rm CE} = 1$ (orange dashed line). The pink and green lines show cases where $\alpha_{\rm CE} = $ 0.3 and 1, and $\alpha_{\rm rec} = $ 0.5 and 0.25 respectively, where $a_f$ just exceeds the 0.15 AU mark.}
    \label{fig:alpha_rec}
\end{figure}

As shown in Figure \ref{fig:alpha_rec}, we find that for a canonical value of $\alpha_{\rm CE} = 0.3$, we require $\alpha_{\rm rec} \gtrsim 0.5$, and for $\alpha_{\rm CE} = 1$, we require $\alpha_{\rm rec} \gtrsim 0.25$ for $a_f \gtrsim 0.15$ AU. Thus, our models point towards a relatively large fraction of recombination energy being needed to produce wide PCEBs. This is not  unexpected, as recombination energy dominates the internal energy of the envelopes of cool stars. For typical stellar compositions, recombination energy dominates over internal energy for temperatures below  $\approx 2 \times 10^5$ K \citep{Ivanova2013A&ARv}. This boundary lies deep in the envelope of our models on the AGB, at roughly 20\% of the stars' radii. 

If a large fraction of recombination energy actually escapes, our results would imply that other sources of energy must be invoked to produce wide PCEBs \citep[e.g. jets from the accreting star;][]{Sabach2017MNRAS, MorenoMendez2017MNRAS}. We defer more detailed calculations and further discussion of this topic to future work. 

\subsection{Models for lower- and higher-mass giants}

We performed similar calculations for a $1\,M_{\odot}$ red giant, which might be a progenitor to a wide PCEB hosting a lower-mass ($0.5-0.6\,M_{\odot}$) WD. We evolved a $1\,M_{\odot}$, solar-metallicity star using inlists from the \texttt{1M\_pre\_ms\_to\_wd} calculation in the MESA test suite. The default wind parameters in that calculation are set so that the mass loss is unusually efficient on the AGB (\texttt{Blocker\_scaling\_factor} = 0.7). This speeds up the calculation by preventing the star from evolving far up the AGB and encountering thermal pulses, but it terminates the AGB phase unrealistically early. We instead set \texttt{Blocker\_scaling\_factor} = 0.05 following \citet{Farmer2015ApJ}.

\begin{figure*}
    \includegraphics[width=0.255\textwidth]{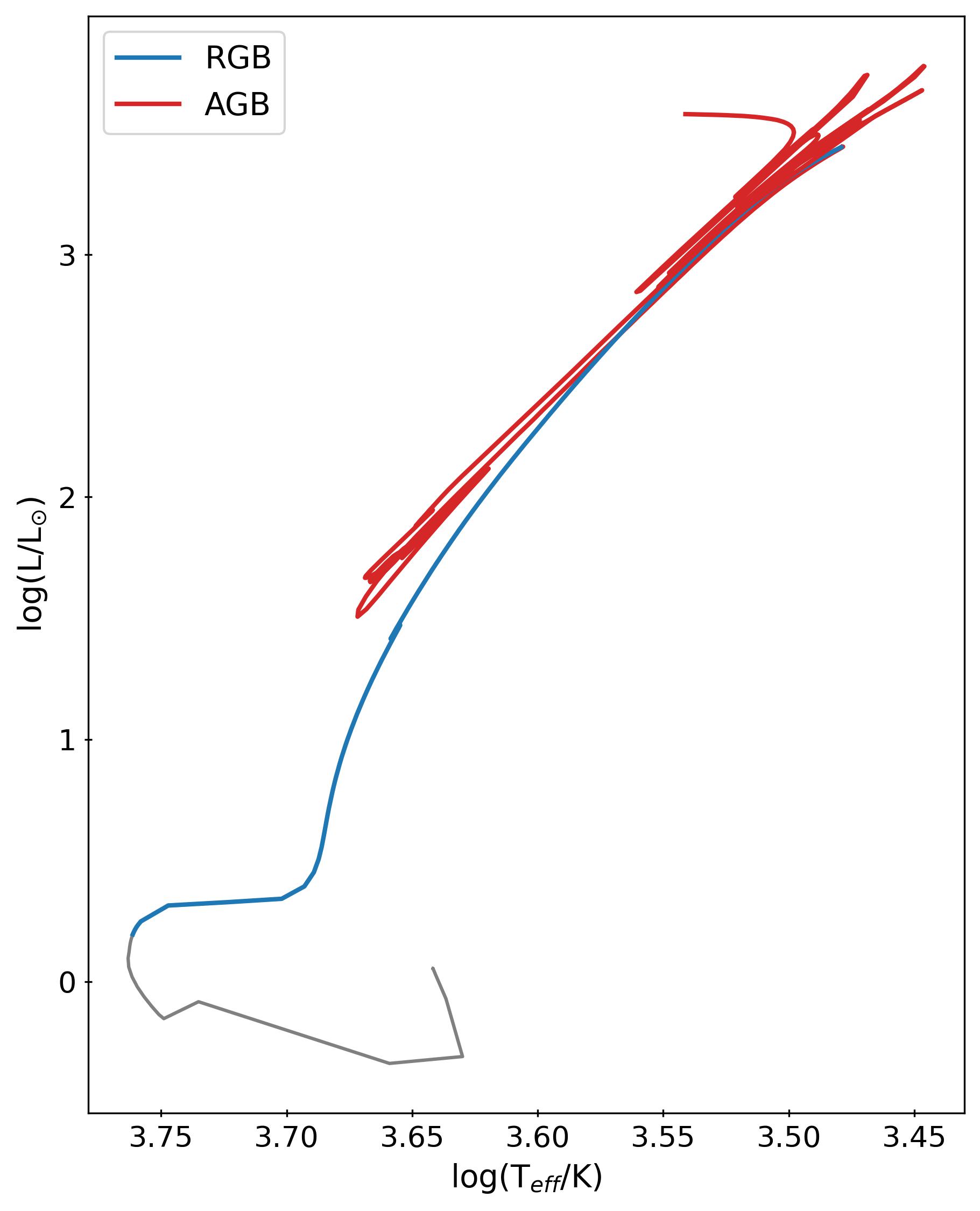}
    \includegraphics[width=0.715\textwidth]{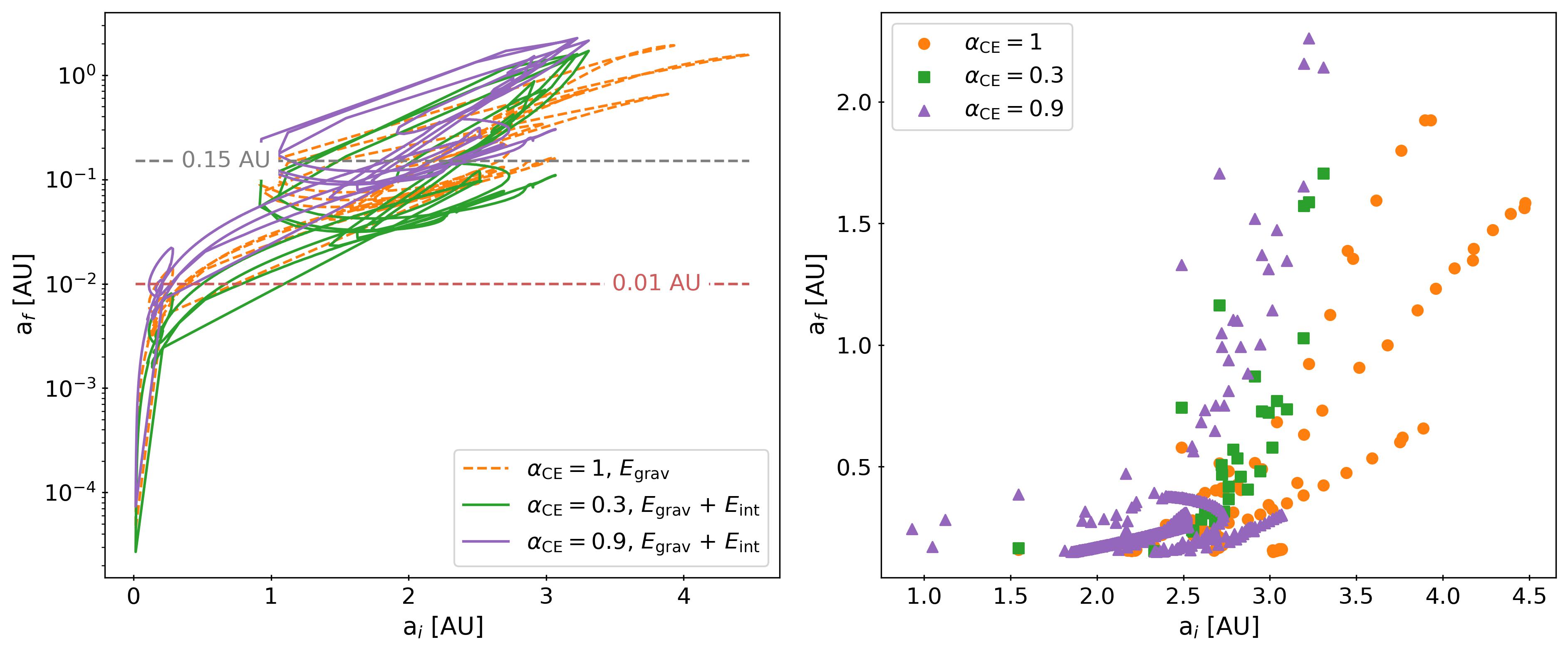}
    \caption{Same as Figure \ref{fig:mesa_af} but for a $1\,M_{\odot}$ model. We omit the $\alpha_{\rm CE} = 0.6$ case to avoid over-crowding. Even in the case where just $E_{\rm grav}$ is considered (orange), there is a range of initial separations for which which wide PCEBs can be produced with $\alpha_{\rm CE}=1$. When $E_{\rm int}$ is included, there is a wide range of initial separations for which wide PCEBs can be produced, even with $\alpha_{\rm CE} < 1$ (green and purple). The models producing wide PCEBs are those in which MT begins during the TP-AGB phase, when the envelope is very weakly bound. }
    \label{fig:mesa_af_1Msun}
\end{figure*}

The same plots as in Figure \ref{fig:mesa_af} but for the $1\,M_{\odot}$ model are shown in Figure \ref{fig:mesa_af_1Msun}. Even in the case where we consider only the gravitational binding energy with $\alpha_{\rm CE} = 1$, there is a range of initial separations ($\sim 1.5 - 4.5$ AU) for which $a_f > 0.15$ AU. This only occurs on the thermally-pulsating phase of the AGB (TP-AGB) where the envelope becomes very loosely bound. Wide separations can also result for much smaller values of $\alpha_{\rm CE} \sim 0.1$ if MT starts at the tip of the AGB. This suggests that it is possible to produce PCEBs in wide orbits containing $0.5-0.6\,M_{\odot}$ WDs (such as the self-lensing binaries) without the need to invoke additional energy sources, but only if the MT begins on the TP-AGB. It should, however, be kept in mind that the TP-AGB phase is a particularly difficult phase to model \citep[e.g.][]{Marigo2007A&A, Girardi2007A&A}, so this conclusion may depend on the adopted stellar models.  

With the inclusion of all of the internal energy, wide PCEB orbits are produced for a broad range of initial separations. The full range of initial separations for which the calculations predict $a_f$ > 0.15 AU is $\sim$ 0.9 to 4.5 AU. For $a_i \sim 3-4.5$ AU, the envelope's binding energy is positive when recombination energy is included, so the $\alpha$ formalism does not straightforwardly predict a final separation. For separations $a_i \lesssim 2.5$ AU, CEE would likely commence on the RGB, preventing the wide PCEB outcome from being realized in practice. Overall, this calculation suggests that efficient envelope ejection can similarly produce wide PCEBs with both low and high mass WDs.

We also ran models of more massive stars, with initial masses of 12 and $20\,M_{\odot}$, that become red supergiants and will leave behind neutron stars or black holes. We find that the envelopes of these stars are significantly more bound than that of the $7\,M_{\odot}$ super-AGB star. This implies that it is difficult to form wide BH/NS + solar-type stars via CEE. Similar conclusions have been reached by other studies \citep[e.g.][]{Kalogera1999ApJ,Kiel2006MNRAS, Giacobbo2018MNRAS, Fragos2019ApJL, El-Badry2023MNRAS}.  

\section{Discussion} \label{sec:discussion}

\subsection{Formation through stable MT?}

Given that the binaries in our sample have wider orbits than traditional PCEBs, it is natural to wonder whether they could have formed through stable MT instead of CEE. We briefly discuss this possibility here.  

The onset of dynamically unstable MT is determined by the donor star's adiabatic response to mass loss. In general, a system in which the donor is more massive than the accretor will tend to be more unstable. The critical mass ratio above which mass transfer is unstable, $q_{\rm crit}$, depends on the stellar structure and evolutionary state of the donor. But in the case of our systems with an ultramassive WD progenitor of mass $\sim 6-7\,M_{\odot}$ and an intermediate MS companion of $\sim 1\,M_{\odot}$, the mass ratio is large enough to exceed even conservative values of $q_{\rm crit} \sim 3-4$ \citep[donor/accretor; e.g.][]{Hjellming1987ApJ, Ge2010ApJ, Temmink2023A&A}. 

The non-zero eccentricities of our observed systems also points towards CEE, which is expected to be less efficient at tidal circularization than stable RLOF, and may actually drive small eccentricities during the plunge-in phase or via torques from circumbinary material \citep[e.g.][]{Ivanova2013A&ARv}. These arguments suggest that stable mass transfer is unlikely to have formed the systems in our sample. 
 
We would be remiss here to not mention the ``$\gamma$''-formalism,  another commonly used prescription of CEE. This was originally invoked to model the formation of double CO WD binaries, which were thought to require a widening of the orbit after the first CE phase, which cannot occur in the $\alpha$-formalism \citep[][]{Nelemans2000A&A, Nelemans2005MNRAS}. The parameter $\gamma$ can be understood as the ratio of the angular momentum lost per mass of ejected material to the average angular momentum per unit mass of the initial binary \citep{Paczynski1976}. While this formalism can produce the wide orbits seen in our systems, it should be emphasized that it was designed precisely for this purpose and does not fundamentally solve  the issues associated with energy conservation, which must still hold. It has also be argued that the $\gamma$-formalism does not actually describe the CEE -- the result of unstable MT -- but instead a phase of stable, non-conservative MT. See Section 5 of \citet{Ivanova2013A&ARv} for further discussion on this formalism.

\subsection{Relative frequency of wide and close PCEBs}

% - Relative space density of wide vs close WD binaries and selection biases that affected the SDSS and Gaia samples

The small number of wide PCEBs discovered so far raises the question of whether they are intrinsically rarer than close PCEBs, or just more difficult to detect. Here we describe the selection biases against wide PCEBs in previous surveys. Given the complex and thus far poorly understood selection function of the {\it Gaia} DR3 binary sample, we do not attempt to infer the space density of wide PCEBs here. Instead, we compare the distances to various samples of PCEBs as a rough diagnostic of their relative frequencies. 

We cross-match the sample of literature PCEBs compiled by \citet[][also shown in our Figure \ref{fig:avMwd}]{Zorotovic2010A&A} to {\it Gaia} DR3 to obtain their parallaxes. We find that the median distance to SDSS PCEBs within that sample is 328 pc, which is significantly farther than the median distance of 108 pc for non-SDSS PCEBs in the sample. This likely reflects the fact that most of the non-SDSS objects were discovered serendipitously from all-sky studies of bright stars, in many cases having been recognized as binaries via photometric variability. In contrast, the SDSS objects were discovered spectroscopically from a parent sample that is deep but only observed a small fraction of all stars. Our targets have distances ranging from 80 to 510 pc, with a median of 308 pc (Table \ref{tab:basic_info}). At 80 pc, J1314+3818 is nearer than any of the SDSS PCEBs.  IK\,Peg, another wide PCEB, is at 46 pc which is nearer than the majority of the close PCEBs in the literature. 

A particularly interesting case to consider is the binary G 203-47 \citep{Delfosse1999A&A}. That system contains a $\approx 0.27\,M_{\odot}$ MS star orbiting a dark companion that is almost certainly a WD in a period of 14.7 days, similar to the wide PCEBs studied here. At a distance of only 7.5\,pc, G 203-47 is one of the 10 nearest known WDs, and probably the nearest PCEB! It is 3 times nearer -- corresponding to a 27 times smaller search volume -- than the nearest short-period PCEB, RR Cae, but has been largely overlooked by works attempting to constrain CE physics with PCEBs. While it is dangerous to draw population-level conclusions from a single object, this strongly suggests that wide PCEBs are quite common.

\subsection{Comparison to other surveys} \label{ssec:other_surveys}

While the SDSS survey for PCEBs \citep{Rebassa-Mansergas2007MNRAS} was highly effective at finding WD + M dwarf PCEBs in tight orbits ($\lesssim 1$ day), it was biased against finding systems like the ones presented here. This is because the sample was selected based on RV variations detected in low-resolution BOSS spectra, which are more easily detected in close binaries with short orbital periods. Furthermore, the SDSS PCEB survey identified candidates by searching for sources with composite spectra in which contributions of both the WD and the MS companion were detectable. This leads to a strong bias in favor of low-mass (M dwarf) main-sequence companions.  

The White Dwarf Binary Pathways Survey conducted a search for WD + AFGK PCEBs. They first selected AFGK MS stars from the RAVE and LAMOST surveys, and then cross-matched them to GALEX, identifying objects with UV excess as candidates for having a WD companion \citep{Parsons2016MNRAS}. From these WD+MS binary candidates, they selected PCEB candidates as those binaries with RV variations detectable in their low-resolution multi-epoch spectra, mainly from LAMOST \citep{Rebassa-Mansergas2017MNRAS}. This also leads to a strong bias in favor of short periods. 

The White Dwarf Binary Pathways Survey did find three binaries with orbital periods of several weeks,  but \citet{Lagos2022MNRAS} concluded that they were likely contaminants. The binaries in question had significant eccentricities \citep[$e = 0.266 - 0.497$; see Table 1 of ][]{Lagos2022MNRAS}, atypical of PCEBs. Based on HST spectra and high contrast imaging, they concluded that at least two are hierarchical triples in which the WD is a distant tertiary.  Our objects would likely not have been found by their search because they have negligible UV excess.

\section{Conclusions} \label{sec:conclusion}

We presented five post-common envelope binaries (PCEBs) containing ultra-massive WD candidates and intermediate mass MS stars with long orbital periods (18 - 49 days). These objects were discovered as part of a broader search for compact object binaries from the {\it Gaia} DR3 NSS catalog. Previous surveys identified PCEBs using a combination of RV variability, photometric variability, and UV excess, which made them biased towards finding PCEBs with M dwarfs in short-period orbits. Systems like the ones presented here pose a potential challenge in simplified models of common envelope evolution (CEE) as their formation requires loosely bound donor envelopes which can be quickly ejected, leaving them in wide orbits with non-zero eccentricities. Our main findings are as follows:

\begin{enumerate}
    \item \textit{Nature of the unseen companions}: The companions are dark objects with masses of $1.2-1.4\,M_{\odot}$ -- more massive than the solar-type stars orbiting them. The simplest explanation is that they are WDs. We consider two possible alternatives: (1) a tight binary containing two $\sim 0.65\,M_{\odot}$ MS stars. In the most pessimistic case, such an inner binary could escape detection in 4 of our 5 targets. However, the near-circular orbits we observe -- which would be a natural consequence of tidal circularization if the companions are WDs -- are not expected in this hierarchical triple scenario. 
    No tight hierarchical triples with outer MS stars and circular outer orbits are known, and very few triples are known with outer periods below 1000 days. (2) a neutron star (NS). This is also unlikely due to the circular orbits of our systems, as NSs are expected to be born with natal kicks that send them to highly eccentric orbits. Given these considerations, we proceed under the assumption that the unseen companions are WDs.
    \item \textit{WD masses}: From RVs (Section \ref{sec:rvs}), we measure orbital solutions and mass functions. Combining this with the masses of the luminous components obtained from SED fitting (Section \ref{ssec:sed}), we obtain minimum WD masses. These range from to $1.244\pm0.027$ to $1.418\pm0.033\,M_{\odot}$, all consistent with masses just below the Chandrasekhar limit. One object, J1314+3818, has a Gaia astrometric solution, which we fit simultaneously with the RVs to constrain the inclination. For this object, we obtain a precise mass of $1.324\pm0.037\,M_{\odot}$. Assuming the dark companions are in fact WDs, they are among the most massive WDs known.
    \item \textit{Comparison to other PCEBs}: Our newly discovered systems have longer periods and host more massive WDs and MS stars than most known PCEBs (Figure \ref{fig:avMwd}). The only similar system previously known is IK\,Peg. However, it is important to note that selection effects of most previous searches strongly favored short-period PCEBs with low-mass MS stars.
    \item \textit{Comparison to MSP + WD binaries}: We find that our objects have large eccentricities relative to the bulk of the MSP+WD binaries (Figure \ref{fig:PvEcc}). However, they have similar eccentricities at fixed orbital period to MSP + CO WD binaries, which likely also formed through CEE.
    \item \textit{Evolutionary models of the WD progenitors}: We ran MESA models of a $7\,M_{\odot}$ MS star up the RGB and AGB (Section \ref{sec:mesa}), following its internal energy (thermal + recombination) and calculating the expected final separation according to the $\alpha$ formalism. We find that there is no point in the evolutionary phase of the star where final separations comparable to those of our objects ($\gtrsim 0.15$ AU) are predicted if internal energy does not aid in unbinding the envelope. In the case where internal energy is included, there is a range in initial separations ($\sim 3.5 - 4.5$ AU) where final separations exceed $\sim 0.15$ AU. For initial separations wider than 4.5 AU, the binding energy of the envelope is positive when CEE begins, such that a range of (wide) final separations are plausible.
    \item \textit{Space density}: We compare distances of literature PCEBs to those of our objects and a few other wide PCEBs. At $\sim 80$ pc, J1314+3818 is nearer than any of the SDSS PCEBs \citep{Zorotovic2010A&A}. The median distance of objects in our sample is comparable to that of all literature PCEBs. The nearest known PCEB, G 203-47, has a period of 15 days, much wider than typical PCEBs in the literature with $P< 1$ day. A detailed estimate of the space density of wide PCEBs will have to wait for a better characterized {\it Gaia} selection function, but these early discoveries suggest that it is comparable to or larger than that of close PCEBs. 

    We also note that several binary populations not typically included in PCEB studies, such as post-AGB binaries and  short-period barium and carbon stars, also contain systems with wide orbits that can most readily be explained by efficient CE ejection on the AGB (see Section~\ref{ssec:related_pops}). For future work, modelling all of these populations together may pave the way to a more complete understanding of mass transfer processes resulting in the formation of close to intermediate WD + MS binaries.

\end{enumerate}

\section*{Acknowledgements}

We would like to thank Matthias Schreiber, Monica Zorotovic, Sterl Phinney, and the anonymous referee for providing detailed feedback that helped to improve this manuscript, and Thomas Tauris for enlightening discussions. We also thank Thomas Masseron and Keith Hawkins for assistance with BACCHUS, as well as Hans-Walter Rix and Eleonora Zari for observing some of our objects under their FEROS programs. Finally, we thank Matthias Kruckow, Shi-Jie Gao, Orsola De Marco, and Lev Yungelson for pointing us to relevant and interesting references.

N.Y. and K.E. were supported in part by NSF grant AST-2307232.

This work presents results from the European Space Agency (ESA) space mission Gaia. Gaia data are being processed by the Gaia Data Processing and Analysis Consortium (DPAC). Funding for the DPAC is provided by national institutions, in particular the institutions participating in the Gaia Multi-Lateral Agreement (MLA). The Gaia mission website is https://www.cosmos.esa.int/gaia. The Gaia Archive website is http://archives.esac.esa.int/gaia.

This paper includes data gathered with the 6.5 meter Magellan Telescopes located at Las Campanas Observatory, Chile.

This work has made use of the VALD database, operated at Uppsala University, the Institute of Astronomy RAS in Moscow, and the University of Vienna.

%%%%%%%%%%%%%%%%%%%%%%%%%%%%%%%%%%%%%%%%%%%%%%%%%%
\section*{Data Availability}

The data underlying this article are available upon reasonable request to the corresponding author.



\bibliographystyle{mnras}

%%%%%%%%%%%%%%%%%%%%%%%%%%%%%%%%%%%%%%%%%%%%%%%%%%

%%%%%%%%%%%%%%%%% APPENDICES %%%%%%%%%%%%%%%%%%%%%

\appendix

\section{Radial velocities} \label{appendix:rvs}

The radial velocities for each of our objects are listed in Tables \ref{tab:rv_2117} to \ref{tab:rv_0107}. 

\begin{table}
    \centering
    \begin{tabular}{c c c}
         HJD & RV [km/s]  & Instrument \\
         \hline
         2459753.9514 & -100.36 $\pm$ 0.04 & TRES \\
         2459814.7098 & -13.62 $\pm$ 0.04 & FEROS \\
         2459817.6290 & 11.59 $\pm$ 0.04 & FEROS \\
         2459818.6918 & 6.58 $\pm$ 0.04 & FEROS \\
         2459819.7567 & -5.42 $\pm$ 0.04 & FEROS \\
         2459820.7507 & -21.86 $\pm$ 0.06 & FEROS \\
         2459821.6934 & -39.98 $\pm$ 0.04 & FEROS \\
         2459822.6943 & -59.91 $\pm$ 0.12 & FEROS \\
         2459823.7031 & -78.14 $\pm$ 0.12 & FEROS \\
         2459824.6181 & -91.25 $\pm$ 0.05 & FEROS \\
         2459825.7693 & -101.13 $\pm$ 0.09 & FEROS \\
         2459826.7605 & -102.44 $\pm$ 0.06 & FEROS \\
         2459827.6655 & -97.77 $\pm$ 0.05 & FEROS \\
         2459828.7288 & -85.67 $\pm$ 0.05 & FEROS \\
         2459889.6855 & 10.91 $\pm$ 0.04 & TRES \\
         2459903.5700 & -27.14 $\pm$ 0.06 & FEROS \\
         2460074.9628 & -83.22 $\pm$ 0.04 & TRES \\
         2460080.8559 & -65.56 $\pm$ 0.05 & FEROS \\
         2460091.9633 & -67.47 $\pm$ 0.06 & TRES \\
         2460093.9142 & -95.82 $\pm$ 0.05 & TRES \\
         2460095.9101 & -101.34 $\pm$ 0.04 & TRES \\
         2460097.9193 & -80.48 $\pm$ 0.03 & TRES \\
         2460100.9396 & -23.21 $\pm$ 0.04 & TRES \\
         2460102.9528 & 5.82 $\pm$ 0.04 & TRES \\
         2460106.9231 & -11.16 $\pm$ 0.04 & TRES \\
         2460107.9399 & -29.31 $\pm$ 0.04 & TRES \\
         2460108.9243 & -48.74 $\pm$ 0.04 & TRES \\
         2460111.8596 & -96.28 $\pm$ 0.04 & FEROS \\
         2460140.7176 & 10.41 $\pm$ 0.06 & FEROS \\
    \end{tabular}
    \caption{RVs for J2117+0332.}
    \label{tab:rv_2117}
\end{table}

\begin{table}
    \centering
    \begin{tabular}{c c c}
         HJD & RV [km/s] & Instrument \\
         \hline
         2459900.0100 & -61.22 $\pm$ 0.04 & TRES \\
         2459912.9834 & -21.44 $\pm$ 0.04 & TRES \\
         2459924.9851 & 3.19 $\pm$ 0.04 & TRES \\
         2459934.9399 & -76.47 $\pm$ 0.02 & TRES \\
         2459951.9817 & 19.79 $\pm$ 0.04 & TRES \\
         2459954.9640 & 15.24 $\pm$ 0.04 & TRES \\
         2459958.9732 & -11.88 $\pm$ 0.03 & TRES \\
         2459963.9729 & -58.62 $\pm$ 0.05 & TRES \\
         2459970.9312 & -73.19 $\pm$ 0.05 & TRES \\
         2459972.9333 & -60.89 $\pm$ 0.04 & TRES \\
         2459974.9553 & -43.56 $\pm$ 0.04 & TRES \\
    \end{tabular}
    \caption{RVs for J1111+5515}
    \label{tab:rv_1111}
\end{table}

\begin{table}
    \centering
    \begin{tabular}{c c c}
         HJD & RV [km/s]  & Instrument \\
         \hline
         2459925.0123 & 9.44 $\pm$ 0.04 & TRES \\
         2459936.0571 & -45.70 $\pm$ 0.04 & TRES \\
         2459951.9913 & 22.62 $\pm$ 0.03 & TRES \\
         2459970.9937 & 5.95 $\pm$ 0.04 & TRES \\
         2459975.0058 & -22.33 $\pm$ 0.03 & TRES \\
         2459979.0157 & -41.33 $\pm$ 0.03 & TRES \\
         2459982.0153 & -45.77 $\pm$ 0.04 & TRES \\
         2459988.0460 & -31.05 $\pm$ 0.03 & TRES \\
         2459991.8859 & -10.68 $\pm$ 0.03 & TRES \\
         2460000.8443 & 39.02 $\pm$ 0.03 & TRES \\
         2460007.9507 & 49.84 $\pm$ 0.03 & TRES \\
    \end{tabular}
    \caption{RVs for J1314+3818}
    \label{tab:rv_1314}
\end{table}

\begin{table}
    \centering
    \begin{tabular}{c c c}
         HJD & RV [km/s] & Instrument \\
         \hline
         2459813.7415 & 24.14 $\pm$ 0.04 & FEROS \\
         2459817.6103 & 48.93 $\pm$ 0.04 & FEROS \\
         2459818.8187 & 55.94 $\pm$ 0.06 & FEROS \\
         2459819.7716 & 61.08 $\pm$ 0.03 & FEROS \\
         2459820.7658 & 65.91 $\pm$ 0.04 & FEROS \\
         2459822.7247 & 73.31 $\pm$ 0.05 & FEROS \\
         2459823.7338 & 76.15 $\pm$ 0.05 & FEROS \\
         2459824.7085 & 77.91 $\pm$ 0.05 & FEROS \\
         2459825.7838 & 78.86 $\pm$ 0.06 & FEROS \\
         2459826.7799 & 78.97 $\pm$ 0.03 & FEROS \\
         2459827.6420 & 78.25 $\pm$ 0.03 & FEROS \\
         2459830.6480 & 70.78 $\pm$ 0.04 & FEROS \\
         2459832.6634 & 61.90 $\pm$ 0.05 & FEROS \\
         2459898.5129 & -11.85 $\pm$ 0.03 & FEROS \\
         2459904.5650 & 15.34 $\pm$ 0.04 & FEROS \\
         2459915.5389 & 75.04 $\pm$ 0.07 & FEROS \\
         2459920.5358 & 77.22 $\pm$ 0.04 & FEROS \\
         2460050.8438 & 63.57 $\pm$ 0.03 & FEROS \\
         2460076.8618 & -11.34 $\pm$ 0.03 & FEROS \\
         2460077.8044 & -13.52 $\pm$ 0.03 & FEROS \\
         2460113.7077 & 36.52 $\pm$ 0.02 & FEROS \\
         2460139.7119 & 44.55 $\pm$ 0.02 & FEROS \\
    \end{tabular}
    \caption{RVs for J2034-5037}
    \label{tab:rv_2034}
\end{table}

\begin{table}
    \centering
    \begin{tabular}{c c c}
         HJD & RV [km/s] & Instrument \\
         \hline
         2459813.9113 & 14.98 $\pm$ 0.03 & FEROS \\
         2459817.7473 & 35.35 $\pm$ 0.04 & FEROS \\
         2459818.8898 & 41.02 $\pm$ 0.03 & FEROS \\
         2459819.8547 & 45.46 $\pm$ 0.03 & FEROS \\
         2459820.7829 & 49.27 $\pm$ 0.04 & FEROS \\
         2459821.7682 & 52.86 $\pm$ 0.03 & FEROS \\
         2459822.7691 & 55.67 $\pm$ 0.07 & FEROS \\
         2459823.7667 & 57.84 $\pm$ 0.07 & FEROS \\
         2459824.8124 & 58.92 $\pm$ 0.04 & FEROS \\
         2459825.8175 & 58.90 $\pm$ 0.06 & FEROS \\
         2459826.8445 & 58.08 $\pm$ 0.03 & FEROS \\
         2459827.8081 & 56.18 $\pm$ 0.04 & FEROS \\
         2459828.7738 & 53.31 $\pm$ 0.03 & FEROS \\
         2459829.7258 & 49.85 $\pm$ 0.04 & FEROS \\
         2459898.7135 & -27.52 $\pm$ 0.03 & FEROS \\
         2459904.6857 & -16.73 $\pm$ 0.03 & FEROS \\
         2459922.5500 & 58.70 $\pm$ 0.03 & FEROS \\
         2460112.8601 & 40.72 $\pm$ 0.03 & FEROS \\
         2460140.8231 & -26.44 $\pm$ 0.03 & FEROS \\
         2460158.8008 & 24.81 $\pm$ 0.03 & FEROS \\
    \end{tabular}
    \caption{RVs for J0107-2827}
    \label{tab:rv_0107}
\end{table}

\section{SED contribution from WDs} \label{appendix:wd_sed}

As a test of whether the WD is contributing to the observed photometry, we generated Koester WD models \citep{Tremblay2009ApJ, Koester2010MmSAI} and added them to the fitted SEDs from Section \ref{ssec:sed}. The photometry with and without this addition was calculated and compared. For the WD models, we set log(g) = 9.0 given the large masses of our WDs, and tested a range of temperatures, $T_{\rm eff, WD} = 10000 \ - \ 60000$ K. We used the \texttt{pyphot}\footnote{https://mfouesneau.github.io/pyphot/} package to calculate the synthetic photometry in different filters. 

\begin{figure}
    \centering
    \includegraphics[width=\columnwidth]{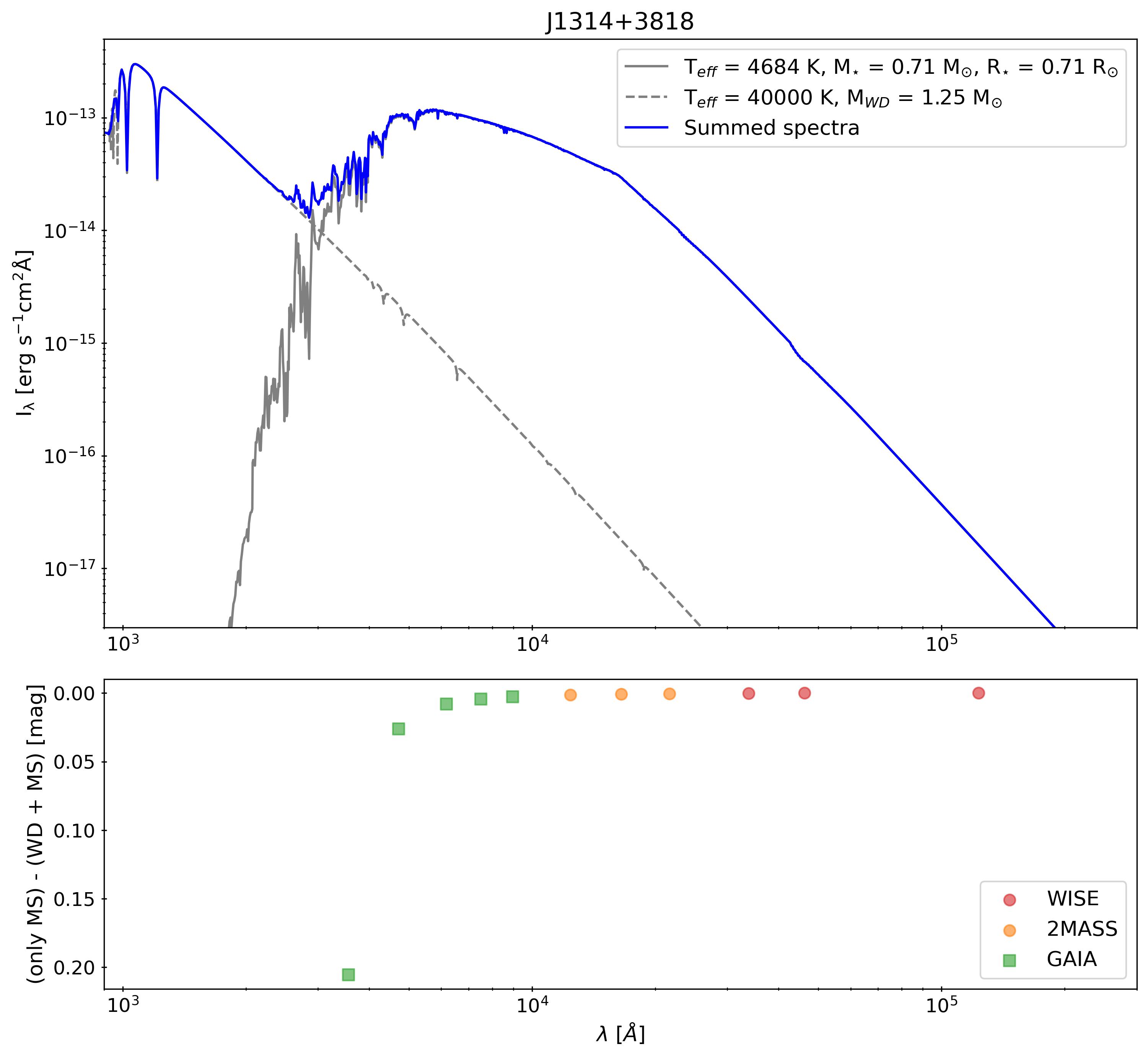}
    \caption{Plot of the fitted SED for J1314+3818 as shown in Figure \ref{fig:seds} (gray solid line), along with the model SED for a WD of $T_{\rm eff, WD} = 40,000$\, (gray dashed lines). The sum of these two SEDs are shown in blue. The lower plot shows the difference in photometry calculated using the gray solid and the blue solid lines. We see that this difference is $\sim 0.2$ mags for the SDSS u band.}
    \label{fig:WD_sed}
\end{figure}

For J1314+3818, which is the most intrinsically faint object amongst the five, we find that the SDSS u band photometry changes by more than 0.2 mags if $T_{\rm eff, WD} \gtrsim 30000$ K. This is large compared to the typical errors on the the photometry which is about 0.02 mags. This is shown on Figure \ref{fig:WD_sed} for $T_{\rm eff, WD} = 40000$ K where the bottom panel shows that the difference is $\sim 0.25$ mags in this case. Therefore, to avoid possible contamination, this point was excluded in our final SED fitting described in Section \ref{ssec:sed}. 

For all other objects, we find no significant contribution, with the SDSS u band photometry changing by less than 0.03 mags for $T_{\rm eff, WD} = 60000$ K. Furthermore, we note that the WD cooling timescale gets longer at lower temperatures meaning that they are more likely to be cool. All this considered, we have included the SDSS u band in our SED fitting for these objects.

We performed the same exercise for the GALEX NUV photometry. In Table \ref{tab:wd_temp_galexnuv}, we summarize the  minimum $T_{\rm eff, WD}$ above which flux contributions of a WD would change the NUV photometry by more than 0.1 mag. We see that WD companions could appreciably change the predicted NUV flux above $T_{\rm eff, WD} \sim$ 7750 K for J1314+3818. The other targets contain MS stars that are brighter in the UV, and so the WDs would be detectable only if they were significantly hotter, with effective temperatures above $T_{\rm eff, WD} \sim $ 19750 - 60000 K. According to the WD cooling models from \citet{Bedard2020ApJ}\footnote{https://www.astro.umontreal.ca/~bergeron/CoolingModels/}, these limits correspond to minimum cooling ages of roughly 3 Gyr for J1314+3818, and 10 - 700 Myr for the other objects. 
 
To avoid potential contamination from the WDs, we exclude the GALEX NUV points in fitting the SEDs. However, as described in Section \ref{ssec:sed}, the observed NUV photometry is nevertheless in good agreement with predictions of the best-fit single star model, placing a lower limit on the ages of the WDs. We note that in Figure \ref{fig:seds}, J1314+3818 may appear to have UV excess, but this owes mostly to the red leak in the GALEX NUV bandpass: calculating synthetic photometry for the best-fit single star model, we find that the excess is only 0.13 mag, which is similar to the uncertainty in the observed photometry of 0.12 mag. If this excess is real, it could be explained by a $1.25\,M_{\odot}$ WD with $T_{\rm eff, WD} \sim$ 7750 K. Such a WD would have an FUV magnitude of 27.3, which is well below the GALEX detection limit. Deeper UV observations (e.g., with the {\it Hubble Space Telescope}) would be needed to assess whether a significant excess is present. Similarly, J0107-2827 has a GALEX excess that is insignificant given the uncertainty.

\begin{table}
    \centering
    \begin{tabular}{c|c}
        \hline
         Name & T$_{\rm eff\, WD, \,min}$ [K] \\
         \hline
         J2117+0332 &  40000 \\
         J1111+5515 &  60000 \\
         J1314+3818 &  7750 \\
         J2034-5037 &  23000 \\
         J0107-2827 &  19750\\
         \hline
    \end{tabular}
    \caption{Minimum WD temperatures above which the WD contribution to the SED would change the GALEX NUV photometry by more than 0.1 mag. WDs cooler than these limits would not be easily detectable in the UV. }
    \label{tab:wd_temp_galexnuv}
\end{table}

%%%%%%%%%%%%%%%%%%%%%%%%%%%%%%%%%%%%%%%%%%%%%%%%%%

% Don't change these lines
\bsp	% typesetting comment
\label{lastpage}
\end{document}